\ifdefined\pdfoutput
  \pdfoutput=1
\fi
\documentclass[aip,reprint,amsmath,amssymb,longbibliography]{revtex4-1}
\usepackage[T1]{fontenc}
\usepackage[utf8]{inputenc}
\usepackage{graphicx}
\usepackage{xcolor}
\usepackage{booktabs}
\usepackage{newtxtext,newtxmath}
\DeclareSymbolFont{operators}{OT1}{ntxtlf}{m}{n}
\SetSymbolFont{operators}{bold}{OT1}{ntxtlf}{b}{n}
\DeclareSymbolFontAlphabet{\mathrm}{operators}
\usepackage{bm}
\usepackage{iftex}
\ifXeTeX
  \usepackage[no-math]{fontspec}
\fi
\usepackage{placeins}

\makeatletter
\def\adjust@abstractwidth{%
  \parindent\z@\relax
  \@totalleftmargin\z@
  \@afterheading\@afterindentfalse
}
\def\frontmatter@@indent{%
  \skip@\@flushglue
  \@flushglue\z@ plus.3\hsize\relax
  \raggedright
  \@totalleftmargin\z@
  \@flushglue\skip@
}
\def\frontmatter@RRAP@format{%
  \addvspace{5\p@}%
  \small
  \raggedright
  \@totalleftmargin\z@
  \everypar{%
    \hbox\bgroup(\@gobble@leavemode@uppercase%)
  }%
  \def\par{%
    \@ifvmode{}{%(
      \unskip)\egroup\@@par
    }%
  }%
}
\makeatother
\begin{document}
\title{Cable construction effects on the lateral forces and wakes
of helically stranded cables under oblique inflow}
% author list: edit front_authors.tex
% ============================================================================================================
% Author list (read by main_reprint.tex before \maketitle). Replace every placeholder below with the real
% information; keep the order of the lines: each \author{...} is followed by its own \affiliation{...}.
%
%  * Names: given name first, family name last, as in your previous publications (e.g. Xiaoming Zhang).
%  * Affiliation: full English name, e.g.
%      \affiliation{School of Aeronautics, Northwestern Polytechnical University, Xi'an 710072, China}
%  * Two authors with the SAME affiliation: give each author its own identical \affiliation{...} line
%    (revtex merges identical affiliations and numbers them automatically).
%  * An author with TWO affiliations: put two \affiliation{...} lines after that \author{...}.
%  * Corresponding author: keep the \email{...} line directly after that author's name; it prints the
%    "Electronic mail" footnote. For two corresponding authors, add an \email{...} after each of them.
%  * The same five names must also be written in sections/declarations.tex (Author Contributions).
% ============================================================================================================
\author{Tielin Ma}
\affiliation{Institute of Unmanned System, Beihang University, Beijing 100191, China}

\author{Junlin Zhang}
\affiliation{School of Aeronautic Science and Engineering, Beihang University, Beijing 100191, China}

\author{Jingcheng Fu}
\email{fujingcheng@buaa.edu.cn}
\affiliation{Institute of Unmanned System, Beihang University, Beijing 100191, China}

\author{Chongwen Jiang}
\affiliation{Institute of Unmanned System, Beihang University, Beijing 100191, China}
\affiliation{School of Aeronautic Science and Engineering, Beihang University, Beijing 100191, China}

\author{Chuan Ye}
\affiliation{Institute of Unmanned System, Beihang University, Beijing 100191, China}

\begin{abstract}
Helically stranded cables in aerial towed systems experience oblique inflow, yet their aeromechanics are conventionally modeled via smooth-cylinder analogies that overlook three-dimensional cross-stream coupling and surface texture. Large-eddy simulations and spectral proper orthogonal decomposition are performed for a smooth cylinder and four stranded cables, all stationary, across four inclinations ($22.5^\circ\text{--}90^\circ$) at $Re \approx 1.0 \times 10^4$. The aerodynamic differences are associated mainly with cross-sectional symmetry order and strand-scale groove depth. Near the wall, force-source decomposition shows that surface pressure accounts for 94.9\%--98.3\% of the mean normal force and dominates the mean lateral force. High-symmetry quasi-circular envelopes maintain balanced circumferential pressure, whereas deep-grooved, low-symmetry tri-lobed geometries induce asymmetric boundary-layer separation that locks in a net lateral pressure difference. In the wake, high-symmetry cables sustain spanwise-synchronized K\'arm\'an shedding under moderate-to-high inclinations, concentrating fluctuation energy into a dominant low-rank mode. Deep grooves instead disrupt shedding coherence, deflect the recirculation zone laterally, and disperse energy across broadband scales. The mean lateral force and the lateral-force fluctuation are therefore controlled by different flow features and decoupled: the quasi-circular $1\times19$ cable carries a minimal mean lateral force but strong periodic fluctuations, whereas the deep-grooved $3\times7$ cable combines weak fluctuations with an appreciable mean lateral force. For cable selection, the results favor quasi-circular, finely textured constructions where a small mean lateral force is required and deep-grooved constructions where lateral-force fluctuations must be limited; the identified mechanisms provide a physical basis for including construction-dependent lateral loads in towing-cable aerodynamic load models.
\end{abstract}
\maketitle
% AIP: all pages numbered (revtex leaves the title page unnumbered)
\thispagestyle{plain}
\fontsize{9pt}{11.5pt}\selectfont
\section{Introduction}
\label{sec:introduction}

\begin{figure}[tb]
 \centering
 \includegraphics[width=\columnwidth]{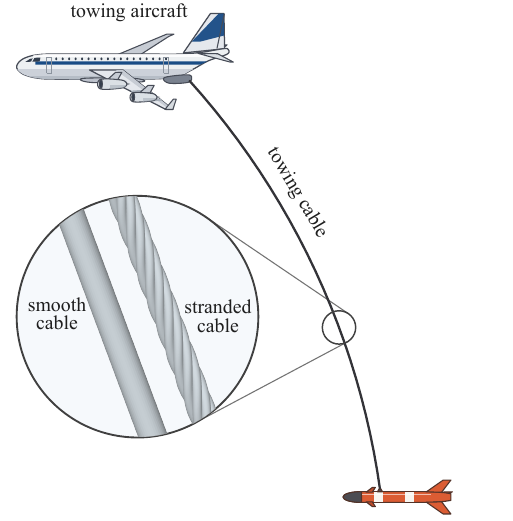}
 \caption{Schematic of an aerial towed system.}
 \label{fig:towed-system}
\end{figure}

An aerial towing system uses a cable as the primary mechanical link to connect an aircraft to a towed payload (Fig.~\ref{fig:towed-system}), with certain designs also transmitting power and signals.\cite{Kakavand2020AeriallyTowed,Dehadrai2021TowedUnderslung} Cables housing optical fibers or electrical conductors are typically protected by elastomeric or rubber jackets; however, mechanically loaded cables are commonly left unjacketed to minimize overall mass, envelope diameter, and bending stiffness. Consequently, their wire strands and helical inter-strand grooves are exposed directly to the airflow, introducing multiscale variations in surface inclination and curvature along both the axial and circumferential directions. Under oblique inflow, this helical surface produces spatially varying fluid--surface interactions that locally alter surface pressure, wall shear, boundary-layer separation and reattachment, and the downstream wake.\cite{Stuart1995GroovedCylinder,Nebres1993YawedCables,Abdelhady2021LES}

Because aerial towing cables frequently extend over several kilometers, minor variations in local aerodynamic loading can accumulate along their length, significantly affecting total aerodynamic drag, tension distribution, and the equilibrium cable configuration.\cite{Liu2026VariableLengthTowing,Wu2022UnderwaterTowed} When predicting towing-system trajectories or local cable tension, treating the cable simply as a smooth cylinder characterized by a single drag coefficient is therefore insufficient. Instead, linking the local flow around the stranded surface to the three-component distributed aerodynamic loads is essential for high-fidelity models of towed decoys and trailing antennas.\cite{Ma2016AerialDecoy,Cheng2023AirborneVLF,Feng2024AerialTowed} Dynamic analyses of long flexible catenary cables and digital-twin frameworks for towed cable--body systems likewise depend on sectional fluid-load models.\cite{Ruan2024CatenaryCable,Westin2025DigitalTwin} Here, surface-pressure redistribution and flow separation determine the pressure force contribution, wall shear provides the viscous contribution, and axial superposition or cancellation determines which local changes survive spanwise integration. Comparable integrated force coefficients can therefore conceal dissimilar surface-loading distributions and wake states, limiting their direct applicability across different cable constructions, inflow angles, and cable lengths.

Although flow past a smooth circular cylinder provides the canonical baseline for bluff-body aerodynamics, even subtle departures from circularity can alter local surface curvature, pressure recovery, and recirculation regions. For instance, at \(Re=10^4\), Wang et al.\ observed both monotonic and nonmonotonic variations in mean surface pressure and wake velocity with increasing roundness imperfections, demonstrating that aerodynamic responses can reverse depending on disturbance amplitude.\cite{Wang2020Roundness} Discrete surface features exhibit a similar sensitivity: spanwise wires and longitudinal fins markedly restructure the near-wake flow and the unsteady loading, and even a single surface groove shifts the separation angle and the balance between pressure and viscous drag according to its location.\cite{Ekmekci2010SpanwiseWire,Islam2020FinnedCylinder,Priyadarsan2023SingleGroove} Moreover, experiments with isolated large-scale surface perturbations demonstrate that boundary-layer separation, local pressure distributions, drag, and vortex shedding frequency depend strongly on perturbation location.\cite{Nebres1993SinglePerturbation}

Surface roughness, including that produced by marine fouling on cables and mooring lines, further perturbs the boundary layer and separated shear layers at geometry-dependent scales.\cite{Maduka2023Biofouling} Across subcritical to transcritical regimes, large-eddy simulations of grooved and rough cylinders have shown that surface topography alters shear-layer transition, separation lines, near-wake topology, Reynolds stresses, and total drag in a regime-dependent manner, and helical grooves can suppress the formation of the K\'arm\'an vortex street.\cite{Cheng2018GroovedCylinder,Rodriguez2017Roughness,Xu2024HelicalGrooves} At lower Reynolds numbers, Kumar and Roy found that the height and spacing of roughness elements govern the formation of local recirculation pockets, induce opposing shifts in pressure and friction drag, and modify both the vortex-formation length and shedding frequency.\cite{Kumar2023Roughness} Overall, these investigations indicate that the aerodynamic impact of surface disturbances depends on their specific geometry, spatial scale, Reynolds number, and inflow conditions. On a stranded cable, these disturbances appear as strand crests and inter-strand grooves repeating along helical paths, giving the cable a three-dimensional geometry that is absent from both uniformly roughened surfaces and bluff bodies with invariant cross sections.

This helical geometry periodically shifts cross-sectional profiles and strand orientations along the cable axis, so that boundary-layer separation, shear-layer development, and wake topology vary with axial phase under identical inflow. For structurally related bluff bodies, such axial variations induce a strongly three-dimensional flow organization; even an abrupt spanwise change in diameter splits the wake into distinct vortex cells connected by vortex dislocations.\cite{Tian2020StepCylinder} The spanwise correlation of cylinder wakes also depends on the aspect ratio or computational span considered.\cite{Norberg1994AspectRatio,Ali2022SpanwiseLength} Numerical simulations of twisted cylinders show that spanwise geometric phase induces three-dimensional separation, delays shear-layer rollup, and produces wavelength locking, oblique vortex shedding, and phase-dependent vortex-formation lengths.\cite{Jung2014TwistedCylinder,Yoon2019TwistedCylinder,Yang2024TwistedWake} Similarly, investigations of helical-straked cylinders found that the helical orientation of the strakes controls Reynolds stress distributions and wake coherence, to the point of breaking up the three-dimensional vortex structures.\cite{Thakurta2025HelicalStrakedWake,Simanto2026HelicalStrakes,ZhaoLee2023HelicalStrake} For flexible cylinders, the strake configuration and yaw angle further govern the suppression of vortex-induced vibration and the resulting fatigue damage.\cite{Xu2017YawStrakes,Quen2018HelicalStrakes,Lu2020YawFatigue} Studies of variable-pitch, wavy, and helically grooved cylinders likewise show that the axial wavelength and groove arrangement modify the near-wake vortex development and aerodynamic loading.\cite{Moon2020VariablePitch,Guo2019WavyCylinder,Xu2024GrooveStartAngle} Although these configurations confirm the aerodynamic significance of axial phase modulation, they differ from stranded cables in cross-sectional topology, boundary conditions, and flow regimes, which prevents direct transfer of their quantitative results. Nevertheless, they identify axial position, helical phase, and repeat wavelength as the physically meaningful parameters for interpreting local separation, wake evolution, and three-component aerodynamic loading on stranded aerial cables.

Beyond the cable's helical geometry, oblique inflow introduces both normal and axial velocity components relative to the cable axis, which alters the velocity composition and spatial development of the local three-dimensional flow. Studies of smooth yawed cylinders provide an essential baseline for isolating this inflow effect from surface topography. Wang, Cheng, and Ting demonstrated that cylinder yaw staggers sectional crossflow forces along the span, causing large local forces to largely cancel in spanwise averages, even though the normal-velocity-normalized Strouhal number remains close to the crossflow value.\cite{Wang2019YawedCylinder} Earlier simulations and wake measurements similarly linked yaw to axial transport, oblique vortex shedding, and reorganized wake vorticity.\cite{Zhao2009YawedCylinder,Zhou2010YawedVorticity,Ramberg1983YawFiniteLength} Subsequent delayed detached-eddy simulations and proper orthogonal decomposition (POD) at \(Re=10^4\) further connected this axial flow to sectional pressure patterns, intermittent lift enhancement, and the convective propagation of force-bearing structures along the span.\cite{Wang2023YawedCylinder} These findings clarify the limits of the independence principle: normal velocity scaling collapses certain frequency characteristics and first-order force statistics, but it does not ensure similarity in local pressure, sectional loading, force fluctuations, or wake phase.\cite{Lou2022YawedSquare} At a fixed freestream speed, changing the cable inclination angle simultaneously alters the normal and tangential velocity components and the normal Reynolds number, so angle-dependent analyses are required to distinguish among these coupled quantities.\cite{Han2012AerialLES,Li2024AngleDependentTowingRopes}

When oblique inflow encounters a helical cable, the geometry-induced axial phase variation interacts with the normal and axial velocity components, generating a nonzero mean lateral force perpendicular to the plane formed by the cable axis and the freestream velocity. This coupled effect has been documented experimentally for various helical cables in yawed flow, and pressure measurements on three-strand twisted ropes attribute the transverse force to the circulation induced by the strand geometry and its orientation to the inflow.\cite{Schacht2022TwistedRopes} In particular, Nebres et al.\ demonstrated that this lateral load arises primarily from asymmetric surface pressure and separation: the side where strands align more closely with the flow experiences delayed boundary-layer separation and lower mean pressure, producing a net transverse force toward that side.\cite{Nebres1993YawedCables} Similarly, in tests motivated by aerial trailing-wire antennas, Stuart et al.\ observed that reversing helix handedness reversed the sign of the groove-induced side-force increment at small inclination angles, with surface tuft alignments confirming a groove-guided, circulation-like mechanism.\cite{Stuart1995GroovedCylinder} Although these investigations, together with side-force measurements on helically grooved slender bodies,\cite{Lua2000HelicalGroove} show that helical surface orientation influences mean lateral loading through asymmetric near-wall flow, they remain limited to isolated configurations. A unified, mechanism-resolved framework is still needed to elucidate how local loading, wake development, and three-component force redistribution systematically evolve with cable construction and inflow inclination.

To bridge this gap, recent investigations have begun resolving the aerodynamic action of stranded and helical surfaces through local surface flows and downstream wake dynamics. In large-eddy simulations and complementary particle image velocimetry (PIV) measurements of \(6\times1\) and \(3\times1\) cables in normal flow, Abdelhady and Wood identified near-stagnant fluid within the inter-strand gaps near the strand crevices, strong axial variations in separation angle, and a deeper base suction than on a circular cylinder. Although the dominant shedding frequency matched that of a circular cylinder, the velocity spectra, Reynolds stresses, and POD modal energy distributions were clearly different.\cite{Abdelhady2021LES,Abdelhady2021PIV}

Wind-tunnel tests on stay cables with helical fillets and wire-mesh coverings similarly showed that localized surface features alter separation lines, transition states, and wake development, with local fluctuating loads partially canceling under spanwise integration.\cite{Christiansen2018PartI,Christiansen2018PartII,Matejicka2022WireMesh} Comparative wind-tunnel and numerical studies across cable surface treatments, cable--wind angles, and cross-sectional shapes further show that the cable geometry significantly modifies the drag, the lift, and their dependence on Reynolds number and inflow angle.\cite{Kleissl2012HelicalFillets,McTavish2020CableSurface,Sang2025UmbilicalCable} These findings demonstrate that a comparable shedding frequency is insufficient to infer similar local loading or wake organization. Although existing studies document isolated variations in local pressure, separation, and wake topology, they have not linked these near-wall flow modifications to the three aerodynamic force components under matched cable geometries and inflow inclinations. For aerial towing cables, the central unresolved question is how these local flow reorganizations are reflected in the normal, lateral, and tangential force components across varying cable constructions and flow angles.

In summary, prior studies show that surface disturbances alter pressure recovery and separation, helical geometry imposes phase-dependent axial flow variations, oblique inflow reshapes wake transport, and the combination of helical geometry and oblique inflow generates mean lateral forces and nonuniform local loading. However, these insights remain fragmented across disparate investigations, and a unified framework linking geometry, inflow, local flow physics, and three-component aerodynamic forces is still lacking for aerial cables. In particular, a construction- and angle-matched comparative analysis is required to determine how normal, lateral, and tangential forces are partitioned into pressure and viscous tractions, and which surface-pressure, separation, and wake signatures accompany this redistribution. Such detailed mechanical insight is essential because integrated aerodynamic coefficients alone cannot explain cable performance: identical global forces can mask markedly different pressure and shear distributions, and local force variations can be diminished or reversed by axial phase cancellation. Local flow reorganization and lengthwise spatial superposition therefore need to be treated together to resolve how near-wall mechanisms determine the integrated aerodynamic loading and to provide the flow-physics basis for assessing the directional aerodynamic forces on aerial towing cables.

Here, large-eddy simulations are employed to systematically investigate the flow physics around a smooth circular cylinder and four representative stranded cables across four inflow inclinations. We evaluate the normal, lateral, and tangential aerodynamic forces across these 20 cases, linking them to pressure/viscous force-source decompositions, surface- and base-pressure recovery, phase-resolved wake fields, and spectral as well as modal diagnostics, the latter having proven effective in resolving coherent structures in turbulent bluff-body wakes and separated flows.\cite{Maleki2025SPODRectangular,Shi2024PlasmaSPOD} This work aims to determine the coupled dependencies among stranded geometry, oblique inflow, local surface flow, and wake organization, thereby clarifying the physical origins of the observed aerodynamic trends. This perspective provides a basis for screening candidate towing-cable constructions according to their directional mean loading and lateral unsteadiness. The remainder of this paper is organized as follows: Sec.~II outlines the computational models, flow parameters, and numerical verification; Sec.~III presents the aerodynamic load balances, surface separation patterns, wake transport dynamics, unsteady spectral coherence, and modal organization; and Sec.~IV summarizes the main conclusions.

\section{Geometry and Numerical Method}
\label{sec:methods}

\begin{figure*}[tp]
 \centering
 \begin{minipage}[b]{0.32\textwidth}
  \centering
  \includegraphics[trim=0 25pt 0 25pt, clip, width=\linewidth]{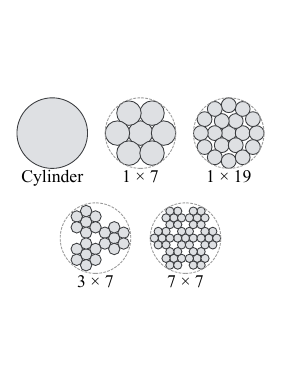}
  \par\vspace{3pt}{\fontsize{8.5pt}{10pt}\selectfont\bfseries (a)}
 \end{minipage}\hfill
 \begin{minipage}[b]{0.32\textwidth}
  \centering
  \includegraphics[width=\linewidth]{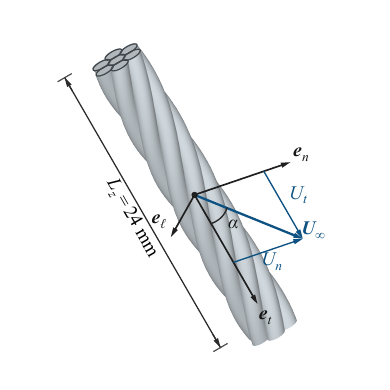}
  \par\vspace{3pt}{\fontsize{8.5pt}{10pt}\selectfont\bfseries (b)}
 \end{minipage}\hfill
 \begin{minipage}[b]{0.32\textwidth}
  \centering
  \includegraphics[width=\linewidth]{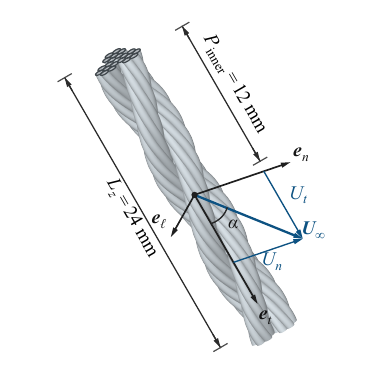}
  \par\vspace{3pt}{\fontsize{8.5pt}{10pt}\selectfont\bfseries (c)}
 \end{minipage}
 \caption{Nominal cable cross sections and representative stranded geometries, with cable-fixed coordinates and inflow velocity components: (a) cross-sectional contours of the five configurations; (b) single-strand $1\times7$ cable (Z lay); and (c) multi-strand $3\times7$ cable (Z/S lay).}
 \label{fig:geometry}
\end{figure*}

\subsection{Geometry and flow-condition definitions}
\label{sec:geometry}

The computational matrix comprises a smooth circular cylinder and four stranded cable configurations, denoted $1\times7$, $1\times19$, $3\times7$, and $7\times7$. Here, $m\times n$ represents a cable formed by $m$ strands, each containing $n$ wires. Each configuration is evaluated at four inflow inclinations, $\alpha=22.5^\circ$, $45^\circ$, $67.5^\circ$, and $90^\circ$, yielding 20 cases in total. All models share an identical envelope diameter, $D=4~\mathrm{mm}$, providing a consistent reference scale for comparing the local flow physics and aerodynamic responses across configurations. In subsequent discussions, ``single-strand'' specifically denotes the $1\times7$ and $1\times19$ configurations, while ``multi-strand'' refers to the $3\times7$ and $7\times7$ cables, thereby distinguishing strand-level architecture from wire-level surface texture.

The axial model length $L_z$ spans one lay pitch. The single-strand and outer-strand lay pitches are set to $6D=24~\mathrm{mm}$, falling within the limits specified by applicable engineering standards.\cite{YBT5197AviationWireRopes,MILDTL83420F} For multi-strand cables, the internal-wire pitch is $P_{\mathrm{inner}}=3D=12~\mathrm{mm}$, which equals half the outer-strand pitch and complies with construction-dependent standards.\cite{MILDTL83420F} Single-strand cables adopt a right-handed (Z) lay, and multi-strand cables combine a right-handed (Z) outer-strand lay with a left-handed (S) internal-wire lay. To facilitate mesh generation and geometrically approximate the contact-induced compaction observed in practical cables, adjacent wires are slightly overlapped and merged. Figure~\ref{fig:geometry} illustrates the nominal cross sections and representative cable geometries.

A cable-fixed right-handed Cartesian coordinate system is defined by the orthonormal basis $(\bm e_n,\bm e_\ell,\bm e_t)=(\bm e_X,\bm e_Y,\bm e_Z)$, where $X$, $Y$, and $Z$ designate the solver Cartesian directions, and the subscripts $n$, $\ell$, and $t$ denote the normal, lateral, and tangential directions, respectively; the coordinates $(x,y,z)$ are measured from the cable axis along $\bm e_n$, $\bm e_\ell$, and $\bm e_t$. The inflow inclination $\alpha$ (equivalent to a yaw angle of $90^\circ-\alpha$ as commonly defined for yawed cylinders) is defined as the angle between the freestream velocity vector $\bm U_\infty$ and the cable axial direction $\bm e_t$, such that $\alpha=90^\circ$ corresponds to pure normal crossflow. The total freestream speed is fixed at $U_\infty=37~\mathrm{m\,s^{-1}}$, which, evaluated with standard sea-level air properties and the envelope diameter $D$, corresponds to a nominal Reynolds number of $Re\approx1.0\times10^4$. With zero lateral velocity prescribed at the inflow boundaries, the incoming freestream velocity is decomposed as
\begin{equation}
 \bm U_\infty=U_n\bm e_n+U_t\bm e_t,\qquad
 U_n=U_\infty\sin\alpha,\qquad U_t=U_\infty\cos\alpha .
 \label{eq:inflow}
\end{equation}
Consequently, varying $\alpha$ changes the normal and tangential velocity components at a fixed total speed, with all subsequent flow quantities and aerodynamic loads referenced to this cable-fixed framework.

Two reference dynamic pressures are distinguished:
\begin{equation}
 \begin{aligned}
 q_{\mathrm{tot}}&=\tfrac12\rho U_\infty^2,\qquad
 q_n=\tfrac12\rho U_n^2,\\
 C_i(q_\chi)&=\frac{F_i}{q_\chi A_{\mathrm{ref}}},
 \qquad A_{\mathrm{ref}}=DL_z,\qquad i\in\{n,\ell,t\}.
 \end{aligned}
 \label{eq:normalization}
\end{equation}
Here, $q_\chi$ with $\chi\in\{\mathrm{tot},n\}$ denotes normalization by the total- or normal-speed dynamic pressure, with $q_\chi$ applied to both force and surface-pressure coefficients. Surface-pressure coefficients are defined as $C_p=(p-p_\infty)/q_\chi$, where $p$ is the time-averaged wall pressure and $p_\infty$ is the time-averaged static pressure at the inlet. Total-speed coefficients describe load variations at fixed flight speed; normal-speed coefficients allow comparison with canonical crossflow scaling. The corresponding Reynolds and Strouhal numbers are
\begin{equation}
 Re=\frac{\rho U_\infty D}{\mu},\qquad
 Re_n=\frac{\rho U_nD}{\mu},\qquad
 St_n=\frac{fD}{U_n},
 \label{eq:dimensionless}
\end{equation}
where $f$ is the frequency; at the dominant vortex shedding frequency, $St_n$ is the shedding Strouhal number, denoted $St_{\mathrm{p}}$ in Sec.~\ref{sec:unsteady-force}, where $St_n$ also serves as the spectral frequency variable. Thus, varying $\alpha$ at fixed $U_\infty$ simultaneously alters both $Re_n$ and $U_t$.

The mean pressure and viscous force components are reconstructed by integrating the time-averaged wall fields over the solid surface $S$:
\begin{equation}
\begin{aligned}
F_{i,p}&=-\int_S p\,
       \hat{\bm n}_{\mathrm{s}}\cdot\bm e_i\,\mathrm{d}S,\qquad
F_{i,\tau}=\int_S
       \bm\tau_{\mathrm{w}}\cdot\bm e_i\,\mathrm{d}S,
\end{aligned}
\label{eq:force}
\end{equation}
where $\hat{\bm n}_{\mathrm{s}}$ is the outward unit normal vector on the solid surface, $p$ is the time-averaged wall pressure defined above, and $\bm\tau_{\mathrm{w}}$ represents the signed viscous wall traction vector; the total force component in Eq.~\eqref{eq:normalization} is $F_i=F_{i,p}+F_{i,\tau}$. Normalizing these two contributions by $q_{\mathrm{tot}}A_{\mathrm{ref}}$ yields the pressure and viscous force coefficients analyzed subsequently:
\begin{equation}
\begin{aligned}
  C_{i,p}&=\frac{F_{i,p}}{q_{\mathrm{tot}}A_{\mathrm{ref}}},\qquad
  C_{i,\tau}=\frac{F_{i,\tau}}{q_{\mathrm{tot}}A_{\mathrm{ref}}}.
\end{aligned}
\label{eq:components}
\end{equation}

Unless otherwise stated, all force coefficients, velocities, and surface pressures reported herein represent time averages, and fluctuation intensities are denoted by the subscript $\mathrm{rms}$ (root mean square). To characterize the helical periodicity of the cables, an axial phase coordinate is defined along the cable axis $\bm e_t$. For each configuration, the phase-resolved cross sections are indexed by the geometric repeat period $P_{\mathrm{rep}}$, which represents the axial distance over which the nominal cross-sectional pattern recurs. The local axial phase $\phi$ is expressed as
\begin{equation}
 \phi=\frac{2\pi z}{P_{\mathrm{rep}}},
 \label{eq:phase}
\end{equation}
where $z$ is the axial coordinate along $\bm e_t$. For the single-strand configurations, the corresponding repeat periods are $P_{\mathrm{rep}}=4~\mathrm{mm}$ for $1\times7$ and $2~\mathrm{mm}$ for $1\times19$, reflecting their discrete rotational symmetries over one lay pitch. For multi-strand cables, different values of $P_{\mathrm{rep}}$ and phase coordinates are evaluated separately for the outer-strand and internal-wire structural scales as needed.

\subsection{Computational domain, mesh, and numerical settings}
\label{sec:numerics}

Three-dimensional large-eddy simulations (LES) are employed to resolve the unsteady flow past the stationary cable geometries. The large, energy-containing turbulent structures are directly resolved on the computational grid, while the subgrid-scale (SGS) stresses are represented by the constant-coefficient Smagorinsky--Lilly model with $C_s=0.1$. This numerical approach resolves the unsteady boundary-layer separation, wake evolution, and aerodynamic load fluctuations required for the present mechanism analysis while avoiding the high computational cost of direct numerical simulation (DNS). Air is modeled as an incompressible Newtonian fluid with density $\rho=1.225~\mathrm{kg\,m^{-3}}$ and dynamic viscosity $\mu=1.7894\times10^{-5}~\mathrm{Pa\,s}$.

\begin{figure}[tb]
 \centering
 \includegraphics[width=\columnwidth]{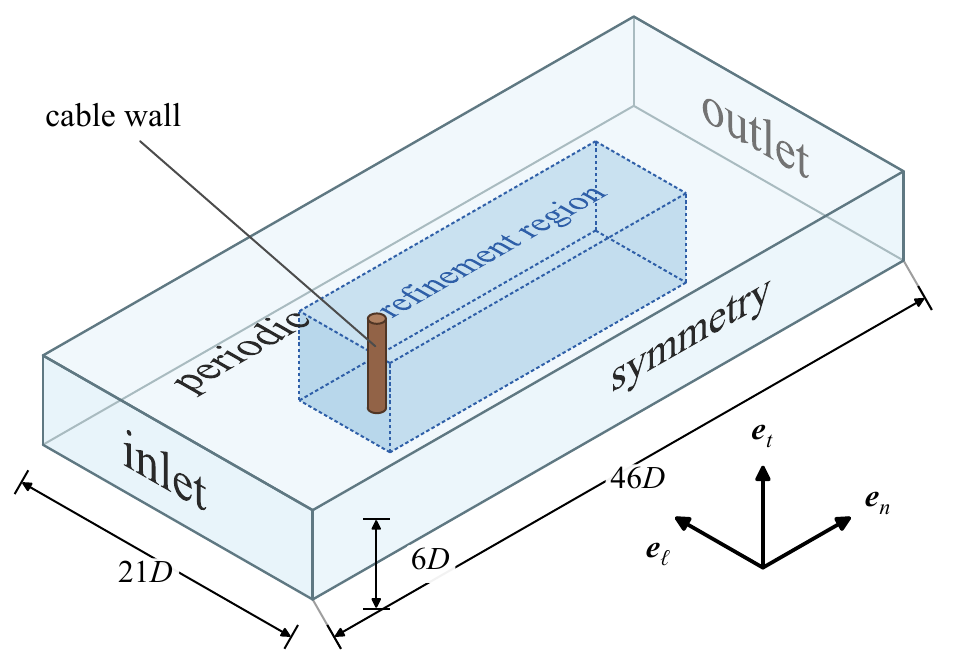}
 \caption{Computational domain, coordinate system, and boundary conditions.}
 \label{fig:domain}
\end{figure}

The computational domain is a rectangular box extending $46D$, $21D$, and $6D$ along the normal ($\bm e_n$), lateral ($\bm e_\ell$), and axial ($\bm e_t$) directions, respectively. Here, the inlet and outlet boundaries are situated $15.5D$ upstream and $30.5D$ downstream of the cable axis, while the lateral boundaries are placed at $10.5D$ on either side. At the inlet, the velocity vector is prescribed using the normal and tangential components defined in Eq.~\eqref{eq:inflow}. A zero-gauge-pressure condition is imposed at the downstream outlet, symmetry boundary conditions are applied to the lateral boundaries, and the stationary cable surface is treated as a no-slip wall. Translational periodic boundary conditions are applied between the two axial end planes. The computational domain, coordinate orientations, and boundary conditions are illustrated in Fig.~\ref{fig:domain}.

\begin{figure*}[tp]
 \centering
 \includegraphics[width=0.9\textwidth]{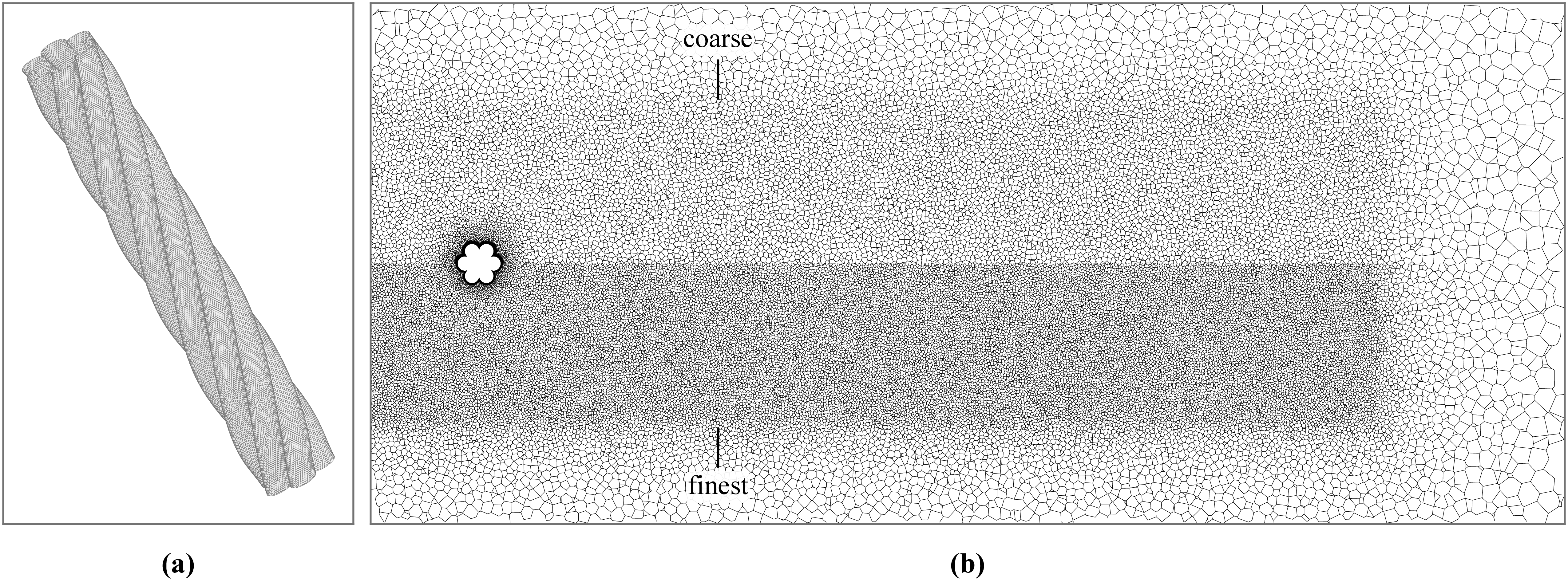}
 \caption{Mesh of the \(1\times7\) cable: (a) surface mesh on the cable wall; and (b) volume-mesh
 cross section in the \(z=0\) plane, with the coarse (upper half) and finest (lower half) meshes.}
 \label{fig:mesh-1x7}
\end{figure*}

Spatial discretization is performed using an unstructured polyhedral mesh with boundary-layer prism layers clustered near the cable wall. Local mesh refinement is applied around the outer wire ridges, inter-wire grooves, and throughout the near-wake region, gradually coarsening toward the far-field boundaries. Four grid levels---designated coarse, medium, fine, and finest---are generated for the grid-independence study. All four grids share the same base settings: a near-wall surface size of $0.025D$, a far-field maximum cell size of $0.5D$, and 15 prism layers with the first-cell height tailored to ensure $y^+\approx0.5$. The grid refinement is controlled solely by the cell size within the local refinement zone, set to $0.125D$, $0.100D$, $0.080D$, and $0.070D$ across the four levels, respectively. Figure~\ref{fig:mesh-1x7} displays the surface mesh of the $1\times7$ cable alongside a cross-sectional view comparing the coarse and finest volume meshes.

The SIMPLE algorithm is used for pressure--velocity coupling. Pressure spatial discretization is handled by a standard second-order scheme, momentum convection is discretized using bounded central differencing, and temporal advancement is achieved via a bounded second-order implicit formulation. The nondimensional time step is defined as
\begin{equation}
\Delta t^*=\frac{\Delta t U_\infty}{D}.
\label{eq:nondim-time-step}
\end{equation}
In normal-velocity units, the time step is $\Delta t U_n/D=\Delta t^*\sin\alpha$. For the time-step sensitivity evaluation at $\alpha=90^\circ$, three nondimensional time steps are compared: $\Delta t^*=0.0023$, $0.0046$, and $0.0092$. After initial transient effects diminish, the production statistics of the mean-flow fields and of the force and probe histories are accumulated over durations of $0.012$--$0.036~\mathrm{s}$, depending on the case, which corresponds to at least 94 convective time units $D/U_n$, or about 19 vortex shedding cycles, in every case.

\subsection{Grid and time-step verification}
\label{sec:verification}

% reference order: the Sec. II C text cites these before the Table I float appears on the page
\nocite{Dong2006NearWake,KravchenkoMoin2000Cylinder,Ali2022SpanwiseLength}
\begin{table*}[tp]
\centering
\caption{Grid and time-step sensitivity and reference results at $90^\circ$ inflow.}
\label{tab:verification}
\fontsize{9pt}{10.8pt}\selectfont
\renewcommand{\arraystretch}{1.0}
\begin{tabular*}{\textwidth}{@{\extracolsep{\fill}}cccccccc@{}}
\toprule
Case & Setting &
\multicolumn{1}{c}{$C_n$} &
\multicolumn{1}{c}{$C_{\ell,\mathrm{rms}}$} &
\multicolumn{1}{c}{$St_n$} &
\multicolumn{1}{c}{$-U_{\min}/U_n$} &
\multicolumn{1}{c}{$L_{\mathrm{r}}/D$} &
\multicolumn{1}{c}{$-C_{p\mathrm{b}}$} \\
\midrule
\multicolumn{8}{@{}l}{\textit{Grid sensitivity}} \\
Cylinder & coarse  & 1.215 & 0.478 & 0.201 & 0.277 & 0.92 & 1.238 \\
         & medium  & 1.135 & 0.324 & 0.203 & 0.322 & 1.03 & 1.117 \\
         & fine    & 1.090 & 0.275 & 0.210 & 0.306 & 1.22 & 1.054 \\
         & finest  & 1.069 & 0.268 & 0.207 & 0.306 & 1.20 & 1.022 \\
\addlinespace[2pt]
$1\times7$ cable & coarse  & 1.122 & 0.219 & 0.198 & 0.294 & 1.14 & 1.053 \\
                 & medium  & 1.117 & 0.212 & 0.206 & 0.299 & 1.12 & 1.032 \\
                 & fine    & 1.119 & 0.200 & 0.205 & 0.294 & 1.18 & 1.047 \\
                 & finest  & 1.078 & 0.186 & 0.206 & 0.279 & 1.24 & 0.980 \\
\addlinespace[2pt]
\multicolumn{8}{@{}l}{\textit{Reference results}} \\
Cylinder & DNS\cite{Dong2006NearWake}
         & 1.143 & -- & 0.203 & 0.249 & 0.82 & 1.129 \\
         & Exp\cite{Dong2006NearWake,Norberg1998LDV,Norberg1987Cylinder}
         & $1.104\text{--}1.186$ & -- & $0.193\text{--}0.202$ & $0.228\text{--}0.380$ & $0.78\text{--}1.02$ & 1.07 \\
$1\times7$ cable & Exp\cite{Richards1962Severn,Batill1989YawedCable}
         & $1.120\text{--}1.360$ & -- & $0.180\text{--}0.220$ & -- & -- & -- \\
         & Exp (PIV)\cite{Abdelhady2021PIV}
         & -- & -- & 0.21 & -- & -- & -- \\
\addlinespace[2pt]
\multicolumn{8}{@{}l}{\textit{Time-step sensitivity on the fine mesh}} \\
Cylinder & $\Delta t^*=0.0023$ & 1.087 & 0.265 & 0.208 & 0.300 & 1.24 & 1.040 \\
         & $\Delta t^*=0.0046$ & 1.090 & 0.275 & 0.210 & 0.306 & 1.22 & 1.054 \\
         & $\Delta t^*=0.0092$ & 1.120 & 0.309 & 0.205 & 0.328 & 1.09 & 1.095 \\
\addlinespace[2pt]
$1\times7$ cable & $\Delta t^*=0.0023$ & 1.101 & 0.193 & 0.207 & 0.282 & 1.19 & 1.012 \\
                 & $\Delta t^*=0.0046$ & 1.119 & 0.200 & 0.205 & 0.294 & 1.18 & 1.047 \\
                 & $\Delta t^*=0.0092$ & 1.083 & 0.172 & 0.207 & 0.323 & 1.30 & 0.992 \\
\bottomrule
\end{tabular*}
\par\smallskip
\begin{minipage}{\textwidth}\footnotesize
Cylinder references are at $Re\approx10^4$: experimental $C_n$ and $St_n$ as compiled by Dong et~al.\cite{Dong2006NearWake}, $U_{\min}/U_n$ and $L_{\mathrm{r}}/D$ from PIV\cite{Dong2006NearWake} and laser Doppler velocimetry (LDV),\cite{Norberg1998LDV} and $-C_{p\mathrm{b}}$ from Norberg\cite{Norberg1987Cylinder}; benchmarks at $Re=1000$--3900 are given in Refs.~\onlinecite{Zukauskas1987,Zdravkovich1990Overview,Abdelhady2019CylinderThermal,Cao2010CylinderShear,Sarwar2020Vortical,Lysenko2012CylinderLES,Henderson1997Nonlinear}. For the $1\times7$ cable, $C_n$ is from overhead conductors,\cite{Richards1962Severn} $St_n$ from wire ropes of other constructions\cite{Batill1989YawedCable} at $Re=5.8\times10^3$--$1.38\times10^4$, and the PIV value\cite{Abdelhady2021PIV} from a six-outer-strand cable at $Re=3100$. Because the present verification runs and the subsequent production runs are separate simulations with targeted adjustments to the statistical averaging duration, some corresponding quantities exhibit minor differences of up to 3\%.
\end{minipage}
\end{table*}

\begin{figure*}[tp]
 \centering
 \includegraphics[width=0.80\textwidth]{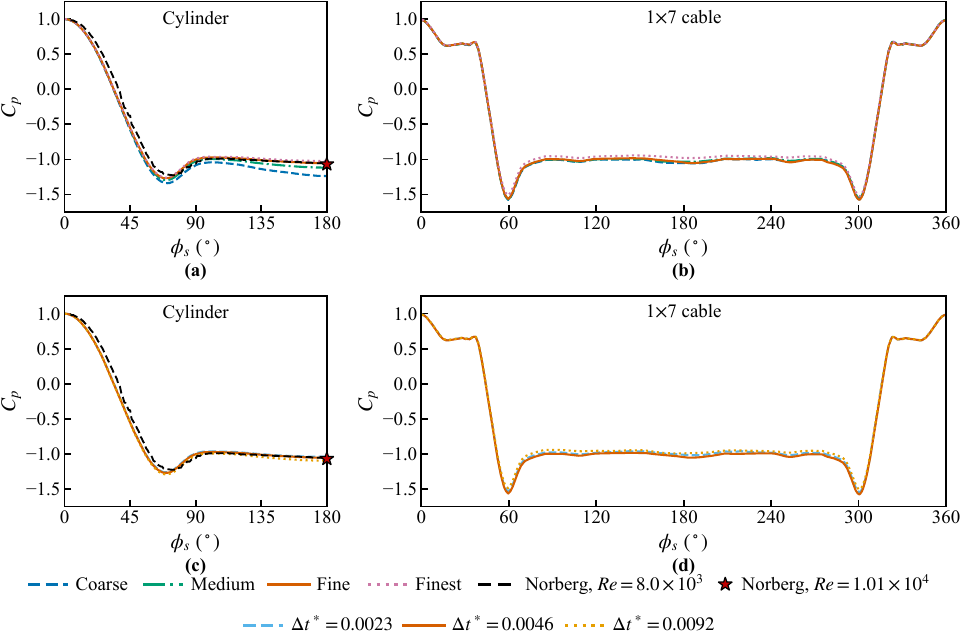}
 \caption{Time-averaged surface-pressure distributions at $\alpha=90^\circ$: (a) smooth cylinder and (b) $1\times7$ cable on the four meshes; (c) smooth cylinder and (d) $1\times7$ cable for the three time steps on the fine mesh. The angle $\phi_s$ is measured from the windward stagnation line. The black dashed line is digitized from Norberg\cite{Norberg1987Cylinder} at $Re=8.0\times10^3$, and the star marks the measured base pressure at $Re=1.01\times10^4$.}
 \label{fig:cp-verification}
\end{figure*}

\looseness=1\relax Grid and time-step verification was conducted for a smooth circular cylinder and a $1\times7$ stranded cable at normal inflow ($\alpha=90^\circ$); the results are summarized in Table~\ref{tab:verification} and Fig.~\ref{fig:cp-verification}. Four mesh levels were evaluated first, followed by three nondimensional time steps on the fine mesh. The assessment considers the mean normal-force coefficient $C_n$, the root-mean-square lateral-force fluctuation $C_{\ell,\mathrm{rms}}$, the normal Strouhal number $St_n$, the normalized minimum centerline velocity $U_{\min}/U_n$, the recirculation length $L_{\mathrm{r}}/D$, and the base pressure coefficient $C_{p\mathrm{b}}$. Here, $U_{\min}$ denotes the minimum normal velocity along the downstream centerline; $L_{\mathrm{r}}/D=x_0/D-0.5$ denotes the wake recirculation length, where $x_0$ is the streamwise position, measured from the cable axis, of the first downstream zero crossing of the time-averaged normal velocity along the centerline, so that the subtracted $0.5D$ refers the length to the nominal rear envelope; and $C_{p\mathrm{b}}$ is the base pressure coefficient, defined as the time-averaged pressure coefficient at the leeward base point. These metrics together characterize the integrated mean loading, the dominant shedding unsteadiness, the wake recirculation, and the leeward base pressure response.

For the smooth cylinder, refining the mesh from coarse to fine reduces $C_n$ from 1.215 to 1.090 and shifts $St_n$ from 0.201 to 0.210, while the further discrepancy between the fine and finest meshes is limited to 1.9\%, 2.6\%, 1.3\%, 0.2\%, 1.6\%, and 3.0\% across $C_n$, $C_{\ell,\mathrm{rms}}$, $St_n$, $U_{\min}/U_n$, $L_{\mathrm{r}}/D$, and $C_{p\mathrm{b}}$, respectively. With the fine-grid solution taken as the reference, the coarse and medium cylinder meshes exhibit relative departures of 11.5\% and 4.1\% in $C_n$, and 4.2\% and 3.2\% in $St_n$, respectively. For the $1\times7$ cable, the corresponding departures in $C_n$ are within 0.3\% and 0.1\%, though $C_{\ell,\mathrm{rms}}$ remains 9.5\% and 6.3\% higher on the coarse and medium grids. The fine-to-finest variations are 3.6\%, 7.0\%, 0.4\%, 4.9\%, 5.1\%, and 6.4\% for $C_n$, $C_{\ell,\mathrm{rms}}$, $St_n$, $U_{\min}/U_n$, $L_{\mathrm{r}}/D$, and $C_{p\mathrm{b}}$, respectively, with the complete grid-convergence data compiled in Table~\ref{tab:verification}.

\looseness=1\relax Against references at $Re\approx10^4$, the fine-grid cylinder $C_n$ lies 1.3\% below the lower experimental bound and 4.6\% below the DNS of Dong et~al.\cite{Dong2006NearWake}, $St_n$ exceeds the measured range by 4.0\%, and $-U_{\min}/U_n$ falls within the experimental interval. The recirculation length is the least well reproduced quantity: the fine-grid value $L_{\mathrm{r}}/D=1.22$ exceeds the measured range of 0.78--1.02, consistent with the known sensitivity of the wake closure point to shear-layer transition and spanwise resolution.\cite{KravchenkoMoin2000Cylinder,Ali2022SpanwiseLength} For the stranded cable, the fine-grid $C_n$ lies within 0.1\% of the lower experimental bound, and $St_n$ lies within the experimental range and 2.4\% below the PIV value for a six-outer-strand cable at $Re=3100$. Consequently, the fine mesh is adopted for all subsequent production simulations.

\looseness=1\relax On the fine mesh, the time-step sensitivity is evaluated using $\Delta t^*=0.0023$, $0.0046$, and $0.0092$. For the smooth cylinder, $\Delta t^*=0.0023$ and $0.0046$ yield essentially identical results for the mean normal force, shedding frequency, minimum centerline velocity, recirculation length, and base pressure coefficient. Increasing the time step to $\Delta t^*=0.0092$ shifts $C_n$ from 1.090 to 1.120, elevates $C_{\ell,\mathrm{rms}}$ from 0.275 to 0.309, and contracts $L_{\mathrm{r}}/D$ from 1.22 to 1.09. For the $1\times7$ stranded cable, while $St_n$ remains between 0.205 and 0.207 for all three settings, the mean force, lateral-force fluctuation, minimum centerline velocity, and recirculation length exhibit their largest departures at $\Delta t^*=0.0092$. Therefore, the fine mesh paired with $\Delta t^*=0.0046$ is selected as the baseline numerical setting for all subsequent production simulations.

\looseness=1\relax Figure~\ref{fig:cp-verification} compares the corresponding mean surface-pressure distributions. For the smooth cylinder, the fine and finest meshes differ by less than 0.02 on the windward half and by at most 0.03 near the base, and $\Delta t^*=0.0023$ stays within 0.02 of the baseline. Against the measurements of Norberg\cite{Norberg1987Cylinder}, the fine-grid leeward plateau agrees to within 0.02 at $Re=8.0\times10^3$, and $-C_{p\mathrm{b}}$ lies within 1.5\% of the value at $Re=1.01\times10^4$, whereas the computed suction peak is about 0.05 deeper and lies about $4^\circ$ farther upstream. For the $1\times7$ cable, the coarse, medium, and fine distributions differ by less than 0.04; the finest mesh and $\Delta t^*=0.0092$ make the leeward plateau about 0.05 less negative.

\section{Results and Discussion}
\label{sec:results}

\subsection{Global three-component aerodynamic loads and pressure--viscous contributions}
\label{sec:global-loads}

\begin{figure*}[tp]
 \centering
 \includegraphics[width=0.85\textwidth]{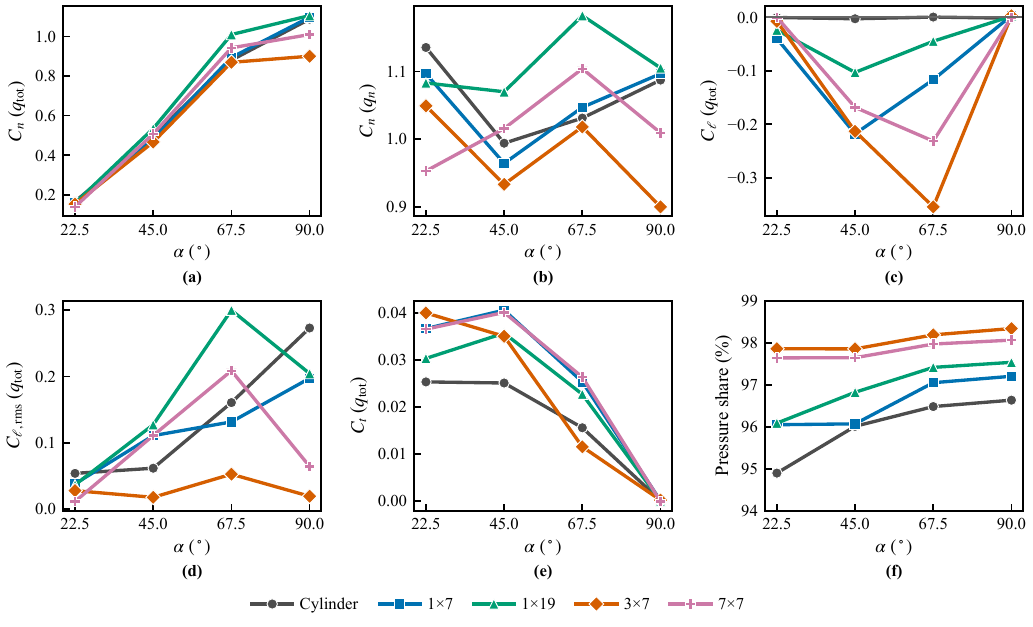}
 \caption{Detailed aerodynamic parameters for all 20 cases: (a) mean normal-force coefficient under total-speed normalization $C_n(q_{\mathrm{tot}})$; (b) mean normal-force coefficient under normal-speed normalization $C_n(q_n)$; (c) mean lateral-force coefficient $C_\ell(q_{\mathrm{tot}})$; (d) root-mean-square lateral-force fluctuation $C_{\ell,\mathrm{rms}}(q_{\mathrm{tot}})$; (e) mean tangential-force coefficient $C_t(q_{\mathrm{tot}})$; and (f) absolute pressure share of the mean normal force.}
 \label{fig:global-load-matrix}
\end{figure*}

We first examine the complete parametric matrix of five geometries and four inflow inclinations to determine how stranded cable surfaces modify the global three-component aerodynamic response. Aerodynamic force coefficients are evaluated under both total-speed ($q_{\mathrm{tot}}$) and normal-speed ($q_n$) normalizations [Eq.~\eqref{eq:normalization}], separating flight-speed load variations from crossflow-scaling departures across geometries. Figure~\ref{fig:global-load-matrix} summarizes the mean normal, lateral, and tangential-force coefficients, the lateral-force fluctuation, and the absolute pressure share of the mean normal force across all 20 cases. Here, the absolute pressure share of component $i$ is defined as $|F_{i,p}|/(|F_{i,p}|+|F_{i,\tau}|)$, which compares the relative magnitudes of pressure and viscous contributions irrespective of algebraic sign.

Under total-speed normalization ($q_{\mathrm{tot}}$), the mean normal-force coefficient $C_n$ increases with inflow inclination $\alpha$ across all five geometries [Fig.~\ref{fig:global-load-matrix}(a)], though both absolute magnitudes and relative configuration rankings evolve. At $\alpha = 22.5^\circ$, $C_n$ spans from $0.140$ for the $7\times7$ cable to $0.166$ for the smooth cylinder; at $\alpha = 90^\circ$, this spread widens to $0.900$--$1.105$. This monotonic increase is associated with the growth of the normal velocity component $U_n$ as $\alpha$ increases at fixed total speed. Furthermore, under normal-speed normalization ($q_n$), the coefficients contract into a much narrower band of approximately $0.90$--$1.18$ [Fig.~\ref{fig:global-load-matrix}(b)], yet they fail to collapse onto a single universal curve: at $\alpha = 67.5^\circ$, the $1\times19$ cable reaches $C_n(q_n) = 1.182$, compared with $1.018$ for the $3\times7$ cable. Normal-speed scaling therefore accounts for the kinematic crossflow contribution and leaves a residual response that depends on the cable surface topology.

Unlike the monotonic behavior of normal loads, the mean lateral-force magnitude $|C_\ell|$ displays a nonmonotonic, configuration-sensitive response to inclination [Fig.~\ref{fig:global-load-matrix}(c)]; at \(\alpha<90^\circ\), the mean lateral force on the stranded cables is directed toward \(-\bm e_\ell\) (\(C_\ell\le0\)), and magnitudes are reported below. In particular, $|C_\ell|$ remains small in absolute terms at $\alpha = 22.5^\circ$, becomes large for several stranded cables at intermediate inclinations ($\alpha = 45^\circ$ and $67.5^\circ$), and returns to near zero under pure crossflow at $\alpha = 90^\circ$. The largest mean lateral force arises in the $3\times7$ cable at $\alpha = 67.5^\circ$, reaching $|C_\ell|(q_{\mathrm{tot}}) = 0.354$, compared with $0.231$ for $7\times7$, $0.117$ for $1\times7$, and $0.045$ for $1\times19$. Already at $\alpha = 45^\circ$, the $1\times7$ and $3\times7$ configurations produce comparable lateral-force magnitudes of $|C_\ell|(q_{\mathrm{tot}}) = 0.218$ and $0.213$, respectively. Throughout the entire inclination range, the smooth cylinder maintains a near-zero mean lateral force ($|C_\ell|(q_{\mathrm{tot}}) < 0.01$), confirming that the nonzero lateral load is an aerodynamic signature of surface stranding.

The inclination corresponding to peak lateral loading differs systematically by cable family: the single-strand cables ($1\times7$ and $1\times19$) peak in magnitude at $\alpha = 45^\circ$, and the multi-strand cables ($3\times7$ and $7\times7$) continue to increase and peak at $\alpha = 67.5^\circ$. Among all stranded geometries, $1\times19$ exhibits the smallest lateral-force magnitude across intermediate inclinations ($|C_\ell|(q_{\mathrm{tot}}) \le 0.103$). This shift in peak inclination largely reflects the normalization: under $q_n$, all stranded cables attain their largest mean lateral-force coefficient near $\alpha = 45^\circ$. Because $|C_\ell|(q_{\mathrm{tot}}) = |C_\ell|(q_n)\sin^2\alpha$ and $\sin^2\alpha$ increases by a factor of 1.71 from $45^\circ$ to $67.5^\circ$, the $q_{\mathrm{tot}}$ peak moves to $67.5^\circ$ when $|C_\ell|(q_n)$ at $67.5^\circ$ retains more than about 59\% of its $45^\circ$ value. This holds for the multi-strand cables, whose $|C_\ell|(q_n)$ decreases by only 3\% ($3\times7$) and 20\% ($7\times7$) between the two inclinations, but not for the single-strand cables, whose $|C_\ell|(q_n)$ decreases by about 70\%. The surface-pressure origin of these lateral-force variations is analyzed in Sec.~\ref{sec:surface-organization}.

The lateral-force fluctuation, measured by $C_{\ell,\mathrm{rms}}$ [Fig.~\ref{fig:global-load-matrix}(d)], displays an independent, configuration-dependent trend across the parameter matrix. At $\alpha = 67.5^\circ$, the $1\times19$ cable develops the strongest unsteady fluctuation ($C_{\ell,\mathrm{rms}}(q_{\mathrm{tot}}) = 0.300$); however, the $3\times7$ cable yields only $0.053$, despite generating the largest mean lateral-force magnitude ($|C_\ell|(q_{\mathrm{tot}}) = 0.354$). Under pure crossflow ($\alpha = 90^\circ$), the mean lateral force returns to near zero across all geometries, yet the fluctuation levels remain clearly configuration-dependent: among the stranded cables, $C_{\ell,\mathrm{rms}}(q_{\mathrm{tot}})$ ranges from $0.019$ for $3\times7$ to $0.204$ for $1\times19$, all below the smooth-cylinder level. The mean lateral-force magnitude $|C_\ell|$ quantifies the stationary asymmetry of the time-averaged flow, and $C_{\ell,\mathrm{rms}}$ reflects the intensity of unsteady vortex shedding fluctuations about this mean state. These two metrics convey complementary dynamic information, and their different combinations separate the aerodynamic responses of the $1\times19$ and $3\times7$ cables.

The mean tangential-force coefficient $C_t$ reflects the three-dimensional flow response under oblique inflow [Fig.~\ref{fig:global-load-matrix}(e)]. Although the axial freestream component produces a baseline viscous tangential load on the smooth cylinder, the departures observed across the stranded cables indicate how the helical surface modifies axial momentum exchange. Across all inclined cases ($\alpha \le 67.5^\circ$), $C_t$ remains nonzero and spans the range of $0.012$--$0.041$, attaining its highest levels at small to intermediate inclinations before returning to near zero at $\alpha = 90^\circ$ under pure crossflow. This variation does not scale monotonically with the nominal number of strands or outer wires: the $3\times7$ cable attains $C_t(q_{\mathrm{tot}}) = 0.040$ at $\alpha = 22.5^\circ$; both the $1\times7$ and $7\times7$ configurations reach peak values of approximately $0.041$ at $\alpha = 45^\circ$.

\begin{figure}[tb]
 \centering
 \includegraphics[width=\columnwidth]{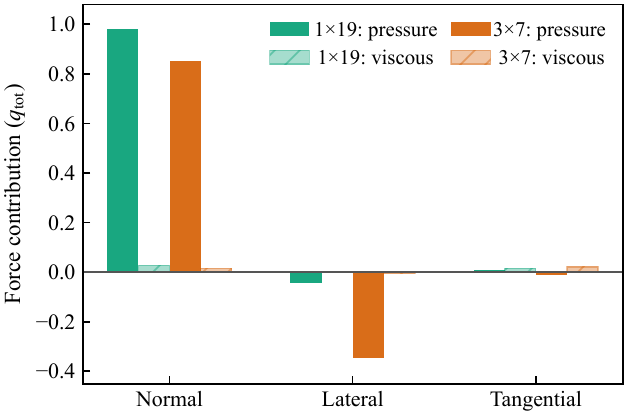}
 \caption{Signed pressure and viscous force contributions across normal, lateral, and tangential directions for the $1\times19$ and $3\times7$ cables at $\alpha = 67.5^\circ$.}
 \label{fig:lateral-bias-source-balance}
\end{figure}

Pressure--viscous force decomposition shows that the mean normal force is dominated by pressure in all 20 cases, with surface pressure accounting for 94.9\%--98.3\% of the total load [Fig.~\ref{fig:global-load-matrix}(f)]. For the mean lateral force, the load is likewise dominated by asymmetric surface pressure. At $\alpha = 67.5^\circ$, for example, the lateral force of the $3\times7$ cable consists of $C_{\ell,p} = -0.347$ and $C_{\ell,\tau} = -0.006$; that of the $1\times19$ cable comprises $C_{\ell,p} = -0.042$ and $C_{\ell,\tau} = -0.003$, confirming that viscous shear represents only a minor fraction of the lateral balance in both configurations. The mean tangential load, however, arises from a delicate balance between viscous friction and surface pressure: at $\alpha = 67.5^\circ$, for $3\times7$, a negative pressure contribution ($C_{t,p} = -0.010$) partially offsets the positive viscous shear ($C_{t,\tau} = 0.021$), resulting in a small net force. For $1\times19$, both contributions act in the same direction ($C_{t,p} = 0.007$ and $C_{t,\tau} = 0.016$). Figure~\ref{fig:lateral-bias-source-balance} summarizes these signed contributions for each force component.

Overall, the integrated load responses across the 20 cases lead to two main observations regarding stranded cable aerodynamics. First, a lower mean normal load does not imply an overall reduction in aerodynamic burden: at $\alpha = 67.5^\circ$, the $3\times7$ cable exhibits a relatively modest normal force but generates the largest mean lateral force across all cases, reflecting a three-dimensional redistribution of the load. The ratio $|C_\ell|/C_n$, which does not depend on the normalization, reaches $0.354/0.869=0.41$ for the $3\times7$ cable, so the resultant of the normal and lateral forces in the cross-sectional plane is deflected by about $22^\circ$ from the plane containing the inflow and the cable axis, compared with $2.6^\circ$ for the $1\times19$ cable ($0.045/1.009$). At $\alpha = 45^\circ$, this deflection exceeds $20^\circ$ for both the $1\times7$ and $3\times7$ cables. A load model that neglects the lateral component would therefore misplace the force direction by an angle of this order.

Second, the mean lateral force is decoupled from the lateral-force fluctuation: the $3\times7$ cable maintains a low lateral-force fluctuation ($C_{\ell,\mathrm{rms}}(q_{\mathrm{tot}}) = 0.053$) together with its pronounced mean lateral force ($|C_\ell|(q_{\mathrm{tot}}) = 0.354$). The $1\times19$ cable, however, develops a strong lateral-force fluctuation ($C_{\ell,\mathrm{rms}}(q_{\mathrm{tot}}) = 0.300$) despite a minimal mean lateral force. These global differences cannot be fully understood from integrated forces alone; they are linked with how helical surface grooves channel the crossflow and reorganize local surface-pressure fields, boundary-layer separation lines, and wake vortex structures. Section~\ref{sec:surface-organization} therefore investigates these local flow fields for representative cable geometries and inflow inclinations.

\subsection{Surface pressure, wall shear, and separation organization}
\label{sec:surface-organization}

\begin{figure*}[tp]
 \centering
 \includegraphics[width=0.85\textwidth]{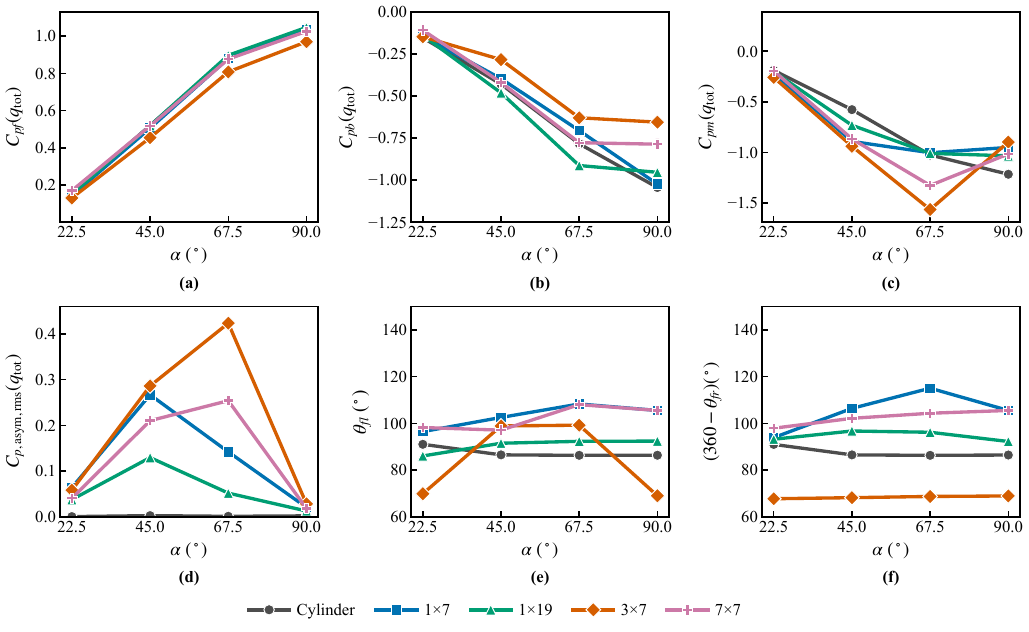}
 \caption{Surface-pressure, wall-shear, and separation-organization indicators across all 20 cases: (a) windward front-point pressure coefficient $C_{p\mathrm{f}}(q_{\mathrm{tot}})$; (b) base pressure coefficient $C_{p\mathrm{b}}(q_{\mathrm{tot}})$; (c) circumferential suction peak (minimum pressure) coefficient $C_{p\mathrm{m}}(q_{\mathrm{tot}})$; (d) circumferential pressure asymmetry metric $C_{p,\mathrm{asym,rms}}(q_{\mathrm{tot}})$; (e) windward-left primary separation angle $\theta_{\mathrm{fl}}$; and (f) reflected windward-right primary separation angle $360^\circ - \theta_{\mathrm{fr}}$.}
 \label{fig:surface-indicators}
\end{figure*}

To clarify the surface origin of the integrated load responses obtained in Sec.~\ref{sec:global-loads}, the time-averaged surface pressure and wall shear stress are evaluated over one lay pitch (\(L_z=6D\)) and mapped onto a unified circumferential coordinate \(\phi_s\) around the nominal outer contour, measured from the windward stagnation line and increasing toward \(+\bm e_\ell\) (distinguished from the axial phase coordinate \(\phi\) defined in Sec.~\ref{sec:methods}). Several diagnostic metrics are used to characterize the key surface features: the windward front-point pressure coefficient \(C_{p\mathrm{f}}\), the circumferential minimum pressure coefficient \(C_{p\mathrm{m}}\), taken as the minimum of the area-weighted, axially averaged distribution \(C_p(\phi_s)\), and the root-mean-square mirror-point pressure difference \(C_{p,\mathrm{asym,rms}}\), which characterizes the circumferential pressure asymmetry,
\begin{equation}
C_{p,\mathrm{asym,rms}}=\Big\{\frac{1}{N_s}\sum_{k=1}^{N_s}\big[C_p(\phi_{s,k})-C_p(360^\circ-\phi_{s,k})\big]^2\Big\}^{1/2},
\label{eq:cp-asym-rms}
\end{equation}
evaluated over the \(N_s=180\) bins of \(1^\circ\) width in \(0^\circ<\phi_s<180^\circ\). Following the kinematic separation criterion of Haller \cite{Haller2004}, applied here to the circumferential component of the time- and axially averaged skin-friction coefficient, boundary-layer separation and reattachment locations are identified from its zero crossings. The circumferential angles of the primary separation points on either side, measured from the windward stagnation line, are denoted by \(\theta_{\mathrm{fl}}\) on the \(+\bm e_\ell\) side (\(0^\circ<\phi_s<180^\circ\)) and \(\theta_{\mathrm{fr}}\) on the \(-\bm e_\ell\) side. Together with \(C_{p\mathrm{b}}\), these indicators summarize the surface-pressure, wall-shear, and separation characteristics across all 20 cases in Fig.~\ref{fig:surface-indicators}.

\begin{figure*}[tp]
\centering
\setlength{\abovecaptionskip}{5pt}
\includegraphics{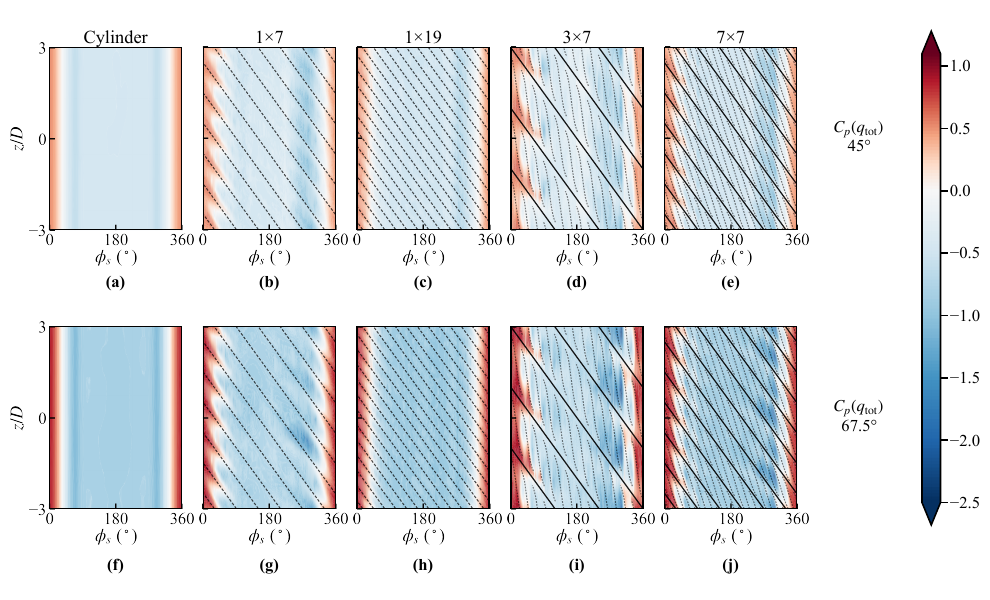}
\caption{Axial phase organization of the mean surface-pressure coefficient \(C_p(q_{\mathrm{tot}})\) across the smooth cylinder and stranded cables at \(\alpha = 45^\circ\) and \(67.5^\circ\). Unwrapped horizontal coordinates \(\phi_s \in [0^\circ, 360^\circ]\) trace the circumferential envelope, and the vertical span \(z/D \in [-3, 3]\) covers one lay pitch (\(L_z=6D\)). Solid and dashed lines mark the inter-strand and inter-wire grooves, respectively.}
\label{fig:surface-phase-maps}
\end{figure*}

Across the parameter space, Figs.~\ref{fig:surface-indicators}(a)--\ref{fig:surface-indicators}(c) summarize the variations in the three primary pressure metrics: \(C_{p\mathrm{f}}\) spans from \(0.104\) to \(1.001\), \(C_{p\mathrm{b}}\) from \(-1.007\) to \(-0.116\), and \(C_{p\mathrm{m}}\) from \(-1.606\) to \(-0.197\). The contrast between the \(1\times19\) and \(3\times7\) cables at \(\alpha = 67.5^\circ\) illustrates this variability. The \(3\times7\) cable exhibits a lower front pressure coefficient (\(C_{p\mathrm{f}}=0.798\)) and a less negative base pressure coefficient (\(C_{p\mathrm{b}}=-0.667\)) than the \(1\times19\) cable (\(C_{p\mathrm{f}}=0.853\) and \(C_{p\mathrm{b}}=-0.962\)). However, its circumferential minimum pressure is much more negative, reaching \(C_{p\mathrm{m}}=-1.606\) compared with \(-1.057\) for \(1\times19\). The deeper suction peak of the \(3\times7\) cable is consistent with local flow acceleration over its prominent outer-strand crests, whereas its less negative base pressure reflects a more modulated leeward recovery (Fig.~\ref{fig:surface-phase-maps}) than the nearly uniform leeward suction of the \(1\times19\) cable.

Across all 20 cases, the suction minimum and the integrated normal load correlate strongly and negatively (\(r = -0.87\)), and the front stagnation pressure correlates almost linearly with the normal load (\(r = 0.99\)). Both correlations, however, largely reflect the common dependence of these quantities on \(\alpha\). At a fixed inclination, the deepest suction does not imply the largest normal load: at both \(\alpha = 45^\circ\) and \(67.5^\circ\), the \(3\times7\) cable combines the deepest suction peak with the lowest normal load among the five configurations. The normal load is therefore determined by the opposing contributions of windward compression and localized suction peaks, so no single scalar pressure indicator is sufficient to represent the integrated aerodynamic force.

A similar contrast between local suction and leeward recovery appears at \(\alpha = 45^\circ\), where the \(3\times7\) and \(1\times19\) cables again form the opposite bounds among the stranded geometries. The \(3\times7\) cable exhibits the least negative base pressure (\(C_{p\mathrm{b}} = -0.310\)) together with the deepest suction peak (\(C_{p\mathrm{m}} = -0.961\)), and the \(1\times19\) cable develops the most negative base pressure (\(C_{p\mathrm{b}} = -0.507\)) with the shallowest suction (\(C_{p\mathrm{m}} = -0.757\)). The \(1\times7\) and \(7\times7\) cables lie strictly between these two bounds, with intermediate base pressure levels (\(C_{p\mathrm{b}} = -0.420\) and \(-0.445\)) and suction minima (\(C_{p\mathrm{m}} = -0.914\) and \(-0.891\)). Figure~\ref{fig:surface-indicators}(b) further shows a systematic dependence on inflow inclination: the cross-configuration spread in \(C_{p\mathrm{b}}\) (defined as the difference between the maximum and minimum values among all configurations at a given inclination) widens monotonically with increasing inclination, expanding from \(0.046\) at \(\alpha = 22.5^\circ\) to \(0.197\) at \(45^\circ\), \(0.295\) at \(67.5^\circ\), and \(0.305\) under pure crossflow (\(\alpha = 90^\circ\)). The leeward base pressure therefore differentiates the cable topologies increasingly as the crossflow component grows, and the geometric sensitivity is largest under pure crossflow.

The circumferential pressure asymmetry shows a systematic sensitivity to both cable configuration and inflow inclination, as evaluated by \(C_{p,\mathrm{asym,rms}}\) in Fig.~\ref{fig:surface-indicators}(d). For the smooth cylinder, \(C_{p,\mathrm{asym,rms}}\) remains bounded between \(5.4\times10^{-4}\) and \(2.6\times10^{-3}\), reflecting time-averaged crossflow reflectional symmetry across all inclinations. The stranded cables, however, develop much larger pressure asymmetry under oblique inflow: the multi-strand \(3\times7\) cable reaches \(0.423\) at \(\alpha = 67.5^\circ\) and \(0.286\) at \(45^\circ\); the \(7\times7\) cable attains \(0.254\) at \(67.5^\circ\). For the single-strand \(1\times19\) cable, the asymmetry peaks instead at a moderate inclination (\(0.130\) at \(45^\circ\)) before decreasing with inclination to \(0.052\) at \(67.5^\circ\) and \(0.013\) at \(90^\circ\). Across the entire 20-case parameter space, \(C_{p,\mathrm{asym,rms}}\) correlates closely with the mean lateral-force magnitude \(|C_\ell|\) (\(r = 0.99\)), with the linear correlation tightening to \(0.997\) at \(\alpha = 45^\circ\) and \(0.998\) at \(67.5^\circ\). Combined with their mutual approach to zero under pure crossflow at \(\alpha = 90^\circ\), this high correlation indicates that the mean lateral force is predominantly driven by oblique-flow-induced surface-pressure redistribution, while asymmetric wall shear stresses exert a weaker influence, consistent with the force decompositions in Sec.~\ref{sec:global-loads}.

The primary separation angles show how stranded surface topography restructures the time-averaged near-wall flow [Figs.~\ref{fig:surface-indicators}(e) and \ref{fig:surface-indicators}(f)]. Here, \(\theta_{\mathrm{fl}}\) denotes the windward-left separation angle, while \(360^\circ - \theta_{\mathrm{fr}}\) represents the reflected windward-right separation angle plotted on the same angular scale from the front stagnation line. For the smooth cylinder, the two separation angles coincide to within \(0.1^\circ\) across all inclinations, remaining bounded between \(86.3^\circ\) and \(91.0^\circ\). On the stranded cables, by contrast, the helical geometry introduces side-to-side differences in separation. For the \(3\times7\) cable, the windward-right separation angle \(360^\circ - \theta_{\mathrm{fr}}\) remains pinned near the first inter-strand valley between \(67.7^\circ\) and \(68.9^\circ\); meanwhile, the windward-left angle \(\theta_{\mathrm{fl}}\) shifts under oblique inflow, registering \(69.9^\circ\), \(98.9^\circ\), \(99.2^\circ\), and \(69.0^\circ\) at \(\alpha = 22.5^\circ\), \(45^\circ\), \(67.5^\circ\), and \(90^\circ\), respectively. For the \(1\times19\) cable, both separation lines remain near \(90^\circ\), with \(\theta_{\mathrm{fl}}\) spanning \(86.0^\circ\)--\(92.4^\circ\) and \(360^\circ - \theta_{\mathrm{fr}}\) spanning \(92.2^\circ\)--\(96.7^\circ\). With increasing inclination, the \(1\times7\) cable yields paired angles \((\theta_{\mathrm{fl}},\, 360^\circ - \theta_{\mathrm{fr}})\) of \(96.4^\circ / 93.9^\circ\), \(102.5^\circ / 106.4^\circ\), \(108.3^\circ / 115.0^\circ\), and \(105.5^\circ / 105.5^\circ\); the corresponding \(7\times7\) pairs evolve through \(98.2^\circ / 97.9^\circ\), \(97.1^\circ / 102.1^\circ\), \(108.0^\circ / 104.3^\circ\), and \(105.5^\circ / 105.5^\circ\). Consequently, both the \(1\times7\) and \(7\times7\) cables converge to the same symmetric separation angle of \(105.5^\circ\) under pure crossflow at \(\alpha = 90^\circ\); the \(3\times7\) cable, however, maintains a pronounced separation asymmetry at \(\alpha = 45^\circ\) and \(67.5^\circ\). This circumferential offset between the primary separation lines accompanies the asymmetric surface-pressure distribution that generates the mean lateral force.

Figure~\ref{fig:surface-phase-maps} presents the axial phase organization of the mean surface pressure in unwrapped two-dimensional representations for all five configurations at \(\alpha = 45^\circ\) and \(67.5^\circ\). In these maps, the horizontal coordinate \(\phi_s\) traces the circumferential surface position, and the vertical axis spans one lay pitch (\(L_z=6D\)). Across the configurations, the surface pressure forms configuration-specific helical patterns whose widths, inclinations, and local patch structures reflect the underlying cable geometry. For the smooth cylinder, the pressure contours remain strictly parallel to the axial direction, displaying complete axial homogeneity. For the single-strand cables (\(1\times7\) and \(1\times19\)), the field is characterized by continuous, inclined helical bands aligned with the outer wire paths. In contrast, the multi-strand cables (\(3\times7\) and \(7\times7\)) exhibit composite two-scale topologies that superimpose fine internal-wire-scale pressure variations onto the broader strand-scale variations, with the tri-lobed envelope of the \(3\times7\) cable producing the strongest spatial modulation. Near the windward stagnation line (\(\phi_s \approx 0^\circ\)), the compression band remains nearly uniform along the axis on the smooth cylinder and the \(1\times19\) cable, whereas the deep inter-strand valleys of the \(3\times7\) cable fragment it into localized pressure dips that shift circumferentially along the axis. When integrated over the body, the alignment between these local pressure patches and the projected surface normal vectors contributes directly to the aerodynamic force balance, consistent with the large mean lateral force and moderate normal load of the \(3\times7\) geometry.

These local surface characteristics link back to the global force balances established in Sec.~\ref{sec:global-loads}. Because the mean normal force is dominated by pressure, variations in windward stagnation compression, suction peak depth, and base recovery determine the net normal resistance. The lateral force, on the other hand, arises from how asymmetric pressure fields are projected through the three-dimensional surface orientation. For the multi-strand cables (\(3\times7\) and \(7\times7\)), the opposite-lay combination of the \(24~\mathrm{mm}\) outer-strand pitch and \(12~\mathrm{mm}\) internal-wire pitch alters the local surface geometry at two different axial length scales. Under oblique inflow, this dual-scale geometric modulation introduces an axial phase dependence into the surface pressure and separation fields. For example, the local separation pinning and suction extrema of the \(3\times7\) cable generate a large mean lateral force together with a moderate normal load. The more symmetric, single-scale surface of the \(1\times19\) cable, however, maintains a minimal mean lateral force but a higher normal resistance. These surface-pressure patterns connect the near-wall topology to the downstream wake dynamics examined in Sec.~\ref{sec:wake-organization}, where the axial-phase-dependent surface structures are linked with the wake reorganization and lateral momentum transport.

\subsection{Mean-wake organization, recirculation, and axial-phase-dependent transport}
\label{sec:wake-organization}

The global force balances in Sec.~\ref{sec:global-loads} and surface flow fields in Sec.~\ref{sec:surface-organization} reveal a marked difference in the dynamic response of the cable families at intermediate inclinations. At \(\alpha = 67.5^\circ\), the \(3\times7\) cable develops a large mean lateral force alongside a weak lateral-force fluctuation (\(\lvert C_\ell\rvert(q_{\mathrm{tot}}) = 0.354\) and \(C_{\ell,\mathrm{rms}}(q_{\mathrm{tot}}) = 0.053\)); the \(1\times19\) cable, however, generates a small mean lateral force along with a strong lateral-force fluctuation (\(\lvert C_\ell\rvert(q_{\mathrm{tot}}) = 0.045\) and \(C_{\ell,\mathrm{rms}}(q_{\mathrm{tot}}) = 0.300\)). To identify the off-body flow features associated with this contrast, we examine how the helical strand--groove topography propagates into the wake, altering the transverse mean-wake displacement, recirculation-zone structure, and axial-phase-dependent momentum transport.

\begin{figure*}[t]
\centering
\setlength{\abovecaptionskip}{5pt}
\includegraphics[width=0.85\textwidth]{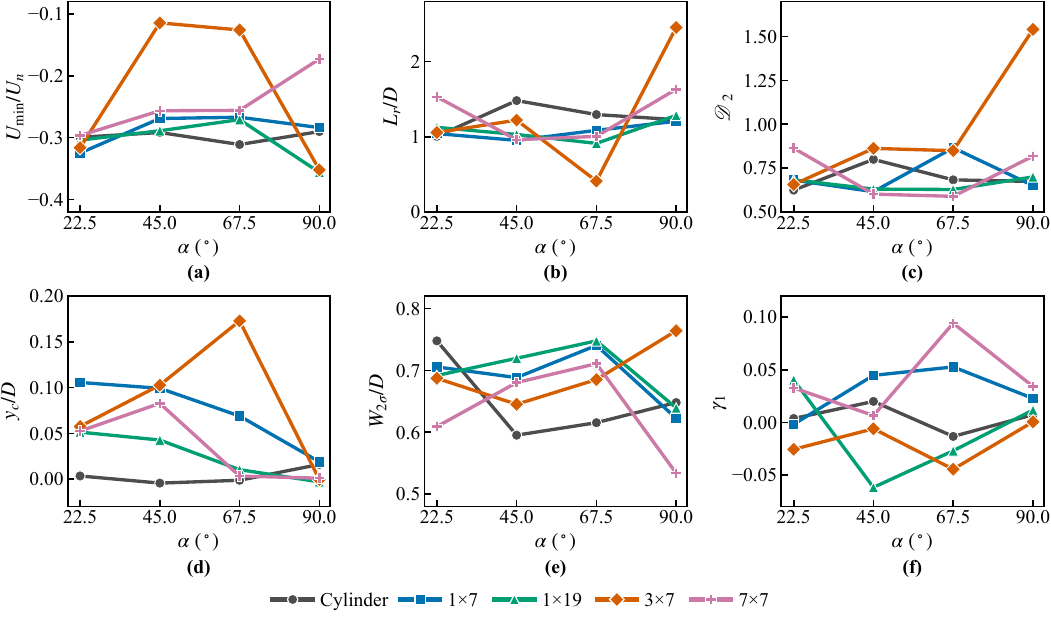}
\caption{Mean-wake indicators across all 20 cases: (a) centerline minimum normal velocity $U_{\min}/U_n$; (b) recirculation length $L_{\mathrm{r}}/D$; (c) integrated sectional velocity deficit $\mathcal{D}_2$ at $x/D = 2$; (d) transverse wake deficit-center displacement $y_{\mathrm{c}}/D$ at $x/D = 2$; (e) wake width $W_{2\sigma}/D$ based on twice the deficit standard deviation at $x/D = 2$; and (f) deficit profile skewness $\gamma_1$ at $x/D = 2$.}
\label{fig:fig9}
\end{figure*}

To characterize the near-body flow structure, the centerline metrics \(U_{\min}/U_n\), \(x_0/D\), and \(L_{\mathrm{r}}/D\) retain the definitions given in Sec.~\ref{sec:verification}. Because oblique inflow and helical surface grooves induce asymmetric wake deflection away from the geometric centerline (\(y = 0\)), relying solely on centerline profiles is insufficient to capture the displaced deficit core. We therefore evaluate the cross-sectional flow field at \(x/D = 2\), introducing a nonnegative velocity deficit weight
\begin{equation}
g(y)=\max\!\left(1-u_n(y)/U_n,\,0\right),
\label{eq:wake-deficit-weight}
\end{equation}
where \(y\) is the lateral coordinate and \(u_n(y)\) is the time-averaged normal velocity along \(x/D = 2\) in the reference section \(z = 0\). The integrated sectional deficit (the subscript 2 denoting \(x/D = 2\)) and deficit-center displacement are evaluated as
\begin{equation}
\mathcal{D}_2=\int g(y)\,\mathrm{d}(y/D),
\label{eq:wake-sectional-deficit}
\end{equation}
\begin{equation}
\frac{y_{\mathrm{c}}}{D}=
\frac{\int (y/D)g(y)\,\mathrm{d}(y/D)}
{\int g(y)\,\mathrm{d}(y/D)}.
\label{eq:wake-center-displacement}
\end{equation}
The effective lateral width and profile skewness are determined from the corresponding higher-order central moments:
\begin{equation}
\left(\frac{\sigma_y}{D}\right)^2=
\frac{\int [(y-y_{\mathrm{c}})/D]^2g(y)\,\mathrm{d}(y/D)}
{\int g(y)\,\mathrm{d}(y/D)},
\label{eq:wake-width-variance}
\end{equation}
\begin{equation}
\frac{W_{2\sigma}}{D}=2\frac{\sigma_y}{D},
\label{eq:wake-width}
\end{equation}
\begin{equation}
\gamma_1=
\frac{\int [(y-y_{\mathrm{c}})/D]^3g(y)\,\mathrm{d}(y/D)}
{\left[\int g(y)\,\mathrm{d}(y/D)\right](\sigma_y/D)^3}.
\label{eq:wake-skewness}
\end{equation}
Here, the integrals are evaluated across the full transverse wake domain using the dimensionless measure \(\mathrm{d}(y/D)\). The four quantities \(\mathcal{D}_2\), \(y_{\mathrm{c}}/D\), \(W_{2\sigma}/D\), and \(\gamma_1\) thus separate variations in deficit magnitude, transverse displacement, lateral spread, and profile asymmetry. Similar deficit-weighted statistical moments have been employed in wake-deflection and rotor-interaction analyses;\cite{Speakman2021MultirotorWake} in the present formulation, they decouple lateral wake deflection from wake spreading and shape distortion.

The \(3\times7\) angle sequence in Fig.~\ref{fig:fig9} indicates that the growth of mean lateral force and surface-pressure asymmetry is reflected in the wake as a transverse displacement of the deficit center. From \(45^\circ\) to \(67.5^\circ\), \(y_{\mathrm{c}}/D\) increases from \(0.103\) to \(0.173\) [Fig.~\ref{fig:fig9}(d)]. Meanwhile, the integrated sectional velocity deficit remains nearly unchanged [\(\mathcal{D}_2 = 0.862\) and \(0.849\), Fig.~\ref{fig:fig9}(c)], and the centerline minimum velocity stays close to \(U_{\min}/U_n \approx -0.12\) [Fig.~\ref{fig:fig9}(a)], while the recirculation length contracts from \(L_{\mathrm{r}}/D = 1.222\) to \(0.410\) [Fig.~\ref{fig:fig9}(b)]. Stronger lateral loading is therefore accompanied primarily by transverse deflection of the deficit core, with little change in the overall sectional velocity deficit. In addition, at the same inclination of \(67.5^\circ\), the \(1\times19\) cable retains a wake that remains close to the centerline (\(y_{\mathrm{c}}/D = 0.010\)) due to its balanced crossflow, consistent with its weaker mean lateral force. This correspondence holds along the \(3\times7\) inclination sequence; the \(7\times7\) cable at \(67.5^\circ\), by contrast, carries a mean lateral force of \(\lvert C_\ell\rvert(q_{\mathrm{tot}}) = 0.231\) with almost no displacement of the deficit center [Fig.~\ref{fig:fig9}(d)].

\begin{figure*}[t]
\centering
\setlength{\abovecaptionskip}{5pt}
\includegraphics[width=0.563\textwidth]{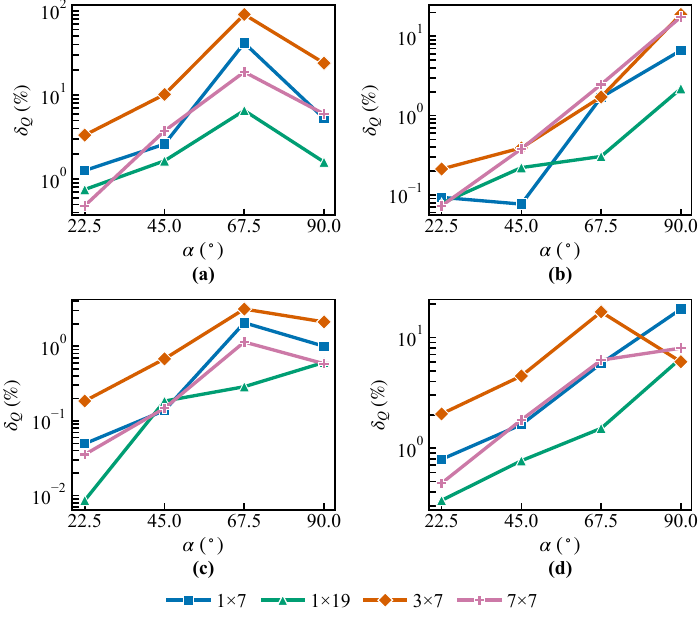}
\caption{Relative range $\delta_Q$ [Eq.~\eqref{eq:relative-range}] across axial sections of four wake indicators against inflow inclination: (a) recirculation length $L_{\mathrm{r}}/D$; (b) velocity deficit integral $I_{\mathrm{def}}$; (c) plane-averaged resolved turbulent kinetic energy $k_{\mathrm{res}}/U_n^2$; and (d) near-wake-averaged lateral velocity magnitude $|u_\ell|/U_n$. Each range is the maximum-minus-minimum difference divided by the absolute value of the arithmetic mean across the axial sections, reported as a percentage.}
\label{fig:fig10}
\end{figure*}

Under pure crossflow, the restored crossflow symmetry eliminates both the mean lateral force and the transverse wake displacement [\(y_{\mathrm{c}}/D = -0.001\), Fig.~\ref{fig:fig9}(d)] for the \(3\times7\) cable. Concurrently, the wake develops an elongated recirculation region: the recirculation length extends to \(L_{\mathrm{r}}/D = 2.455\) [Fig.~\ref{fig:fig9}(b)], the centerline reverse flow deepens to \(U_{\min}/U_n = -0.352\) [Fig.~\ref{fig:fig9}(a)], and the integrated deficit and width increase to \(\mathcal{D}_2 = 1.541\) and \(W_{2\sigma}/D = 0.764\) [Figs.~\ref{fig:fig9}(c) and \ref{fig:fig9}(e)]. These variations reflect two wake states: oblique inflow steers the deficit transversely and shortens the recirculation length, whereas pure crossflow restores symmetry within a deeper, broader, and slower-recovering wake.

To evaluate the cumulative velocity deficit across the wake and distinguish localized profile reorganization from changes in total velocity loss, we define a planar integrated velocity deficit
\begin{equation}
I_{\mathrm{def}}=
\int_{\mathcal A_{\mathrm{def}}}\max\!\left(1-u_n/U_n,0\right)
\frac{\mathrm{d}A}{D^2},
\label{eq:wake-plane-deficit}
\end{equation}
where \(\mathcal A_{\mathrm{def}}\) denotes the sampling domain spanning \(0.5\le x/D\le12\) and \(\lvert y\rvert/D\le4\). Resolved directional turbulent fluctuations are quantified by the Reynolds stress tensor
\begin{equation}
R_{ij}=\langle u_i u_j\rangle-\langle u_i\rangle\langle u_j\rangle,
\label{eq:wake-reynolds-stress}
\end{equation}
where \(\langle\cdot\rangle\) denotes time averaging, \(u_i\) is the instantaneous resolved velocity (elsewhere, \(u_n\) and \(u_\ell\) denote time-averaged components), and \(i,j\in\{n,\ell,t\}\) represent the normal, lateral, and cable-tangential directions in the cable-fixed frame, respectively. The resolved turbulent kinetic energy per unit mass is then evaluated as
\begin{equation}
k_{\mathrm{res}}=\frac{R_{nn}+R_{\ell\ell}+R_{tt}}{2}.
\label{eq:wake-resolved-energy}
\end{equation}
For any flow-field quantity \(Q\), its area-weighted spatial average over a designated planar domain \(\mathcal A\) is computed as
\begin{equation}
\langle Q\rangle_{\mathcal A}=
\frac{1}{A_{\mathcal A}}\int_{\mathcal A} Q\,\mathrm{d}A,
\label{eq:wake-area-average}
\end{equation}
where \(A_{\mathcal A} = \int_{\mathcal A}\mathrm{d}A\) is the area of the domain. For the retained stress and energy summaries, \(\mathcal A\) is the entire fluid cross section of the computational domain (\(46D\times21D\)); these plane-averaged values are reported without additional notation and are identified as plane-averaged in the text and figure captions. The near-wake mean-flow indicator is the near-wake-averaged lateral velocity magnitude \(\lvert u_\ell\rvert/U_n\), evaluated over \(0.5\le x/D\le4\) and \(\lvert y\rvert/D\le4\), with the absolute value taken before area averaging. Both \(R_{ij}\) and \(k_{\mathrm{res}}\) have units of velocity squared and are normalized by \(U_n^2\). The velocity covariance and its half-trace follow the standard Reynolds decomposition,\cite{Pope2000TurbulentFlows} applied here to the resolved velocity field.

\begin{figure*}[tp]
\centering
\setlength{\abovecaptionskip}{5pt}
\includegraphics[width=0.85\textwidth]{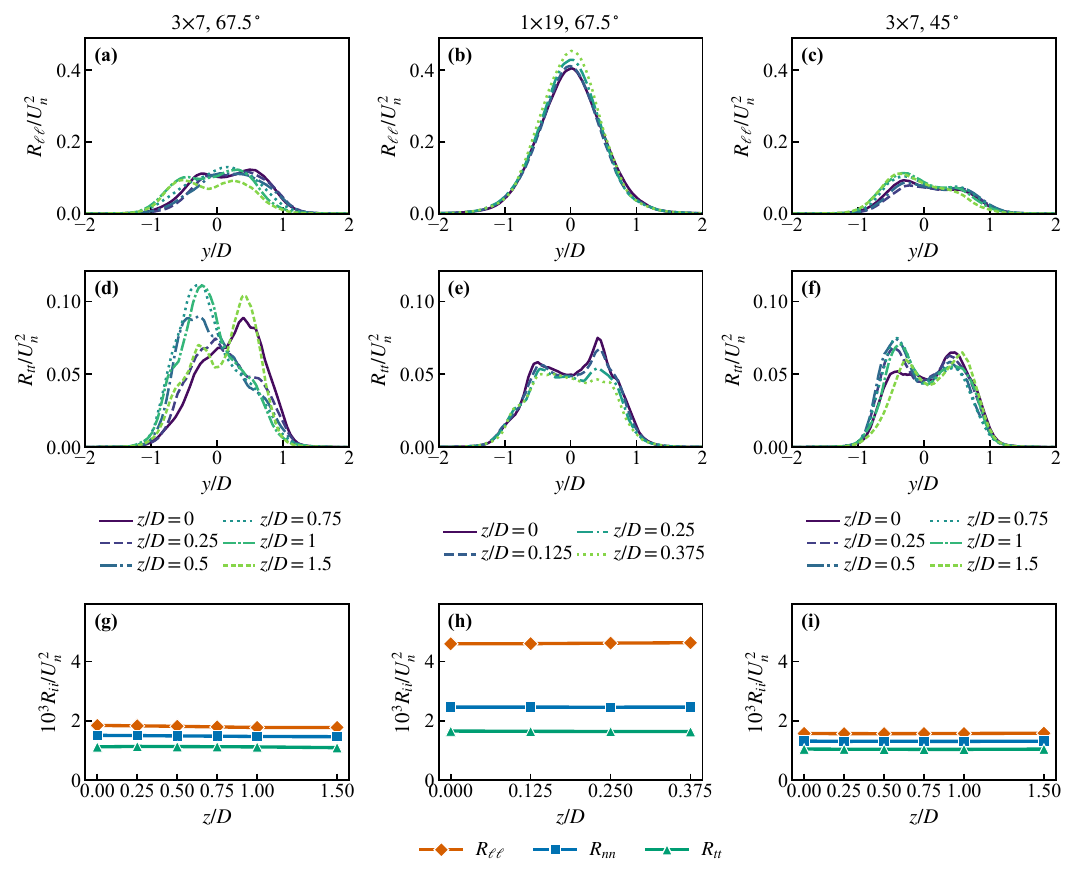}
\caption{Directional velocity fluctuations and area-weighted turbulent stresses across representative stranded cables and inclinations: (a)--(c) local lateral velocity variance \(R_{\ell\ell}/U_n^2\) at \(x/D=2\); (d)--(f) local tangential velocity variance \(R_{tt}/U_n^2\) at \(x/D=2\); and (g)--(i) plane-averaged diagonal components \(10^3R_{ii}/U_n^2\) (\(i=n,\ell,t\); no summation) across axial sections.}
\label{fig:fig11}
\end{figure*}

To quantify the axial flow modulation introduced by helical stranding, Fig.~\ref{fig:fig10} shows the relative range across axial sections for each scalar wake indicator $Q_p$ ($L_{\mathrm{r}}/D$, $I_{\mathrm{def}}$, $k_{\mathrm{res}}/U_n^2$, or $|u_\ell|/U_n$), defined as
\begin{equation}
\delta_Q=\frac{\max_p Q_p-\min_p Q_p}{\left|N_p^{-1}\sum_{p=1}^{N_p} Q_p\right|}\times100\%,
\label{eq:relative-range}
\end{equation}
where $p$ indexes the axial sampling planes ($N_p = 6$ for the $3\times7$ cable and $N_p = 4$ for all other stranded configurations). By normalizing the maximum section-to-section difference by the absolute value of the section-averaged mean, this metric measures the sensitivity of each wake feature to geometric phase.

\begin{figure*}[tp]
\centering
\setlength{\abovecaptionskip}{5pt}
\includegraphics[width=0.85\textwidth]{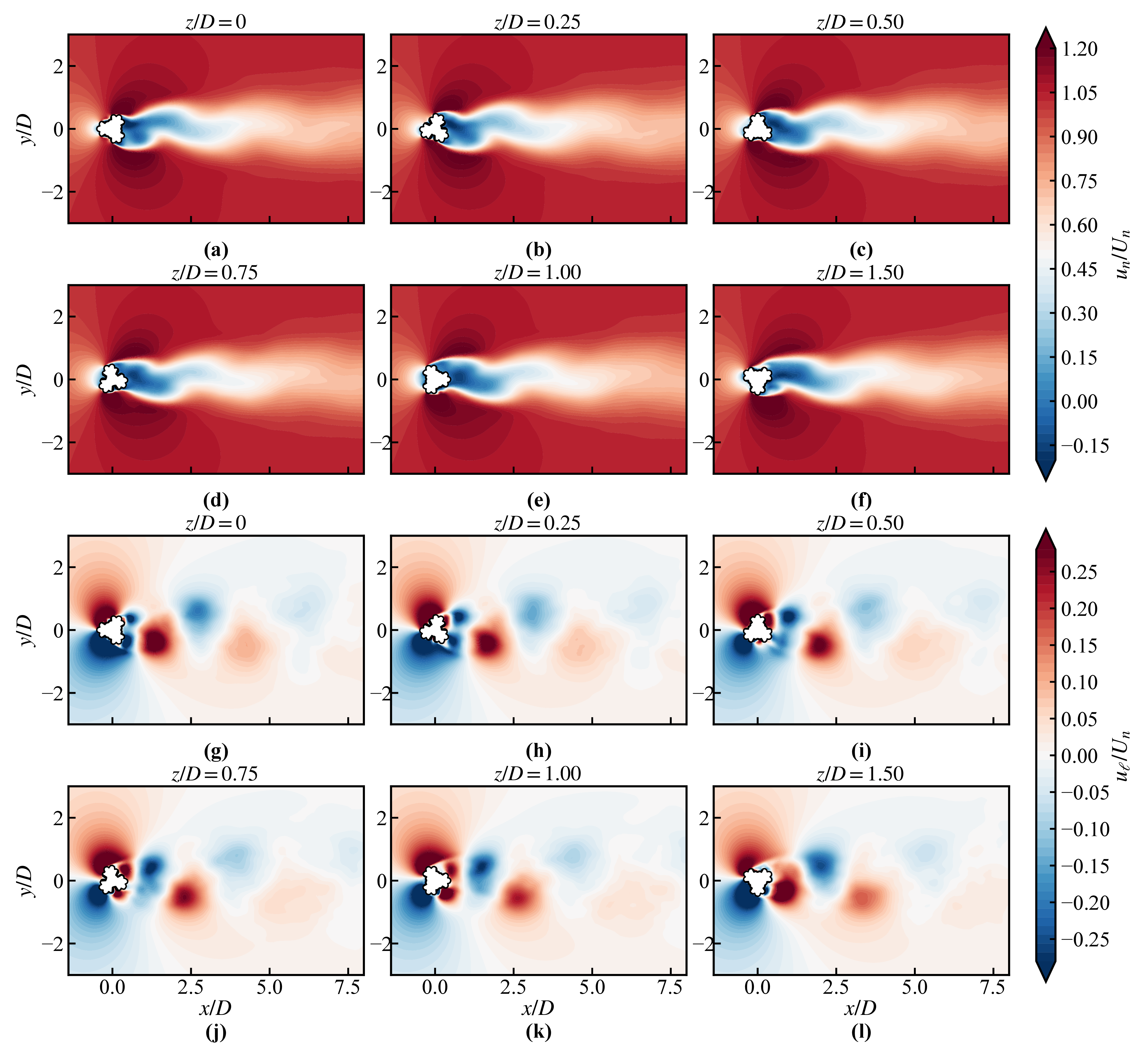}
\caption{Cross-sectional distributions of the mean velocity field across six axial phase sections of the \(3\times7\) stranded cable at \(\alpha = 67.5^\circ\): (a)--(f) normalized normal velocity \(u_n/U_n\) and (g)--(l) normalized lateral velocity \(u_\ell/U_n\). Subplot titles indicate the normalized axial position \(z/D\).}
\label{fig:fig12}
\end{figure*}

At $\alpha = 67.5^\circ$, the $3\times7$ cable exhibits strong axial variations in local recirculation and transverse flow: the relative range reaches $90.8\%$ for the recirculation length $L_{\mathrm{r}}/D$ and $17.0\%$ for the near-wake-averaged lateral velocity $|u_\ell|/U_n$, but the planar deficit integral $I_{\mathrm{def}}$ and the plane-averaged resolved turbulent kinetic energy $k_{\mathrm{res}}/U_n^2$ vary by only $1.72\%$ and $3.14\%$, respectively. The recirculation closure point shifts between $x_0/D = 0.920$ and $1.574$, spanning a streamwise distance of $0.654D$ across the axial sections. At the same inclination, the $1\times19$ cable maintains uniform wake profiles along the span, with relative ranges of $6.56\%$ for $L_{\mathrm{r}}/D$ and $1.51\%$ for the lateral velocity. This contrast indicates that the large mean lateral force on the $3\times7$ cable is accompanied by a local distortion of the recirculation zone along the span, with little change in the overall wake velocity loss. The $7\times7$ cable under pure crossflow ($\alpha = 90^\circ$) shows the opposite pattern: the recirculation length remains relatively stable ($5.96\%$), while the integrated deficit varies by $17.5\%$. In that case, the helical stranding manifests mainly as a section-to-section variation of the integrated deficit, with the streamwise location of the recirculation closure point remaining comparatively stable.

\begin{figure*}[tp]
\centering
\setlength{\abovecaptionskip}{5pt}
\includegraphics[width=0.90\textwidth]{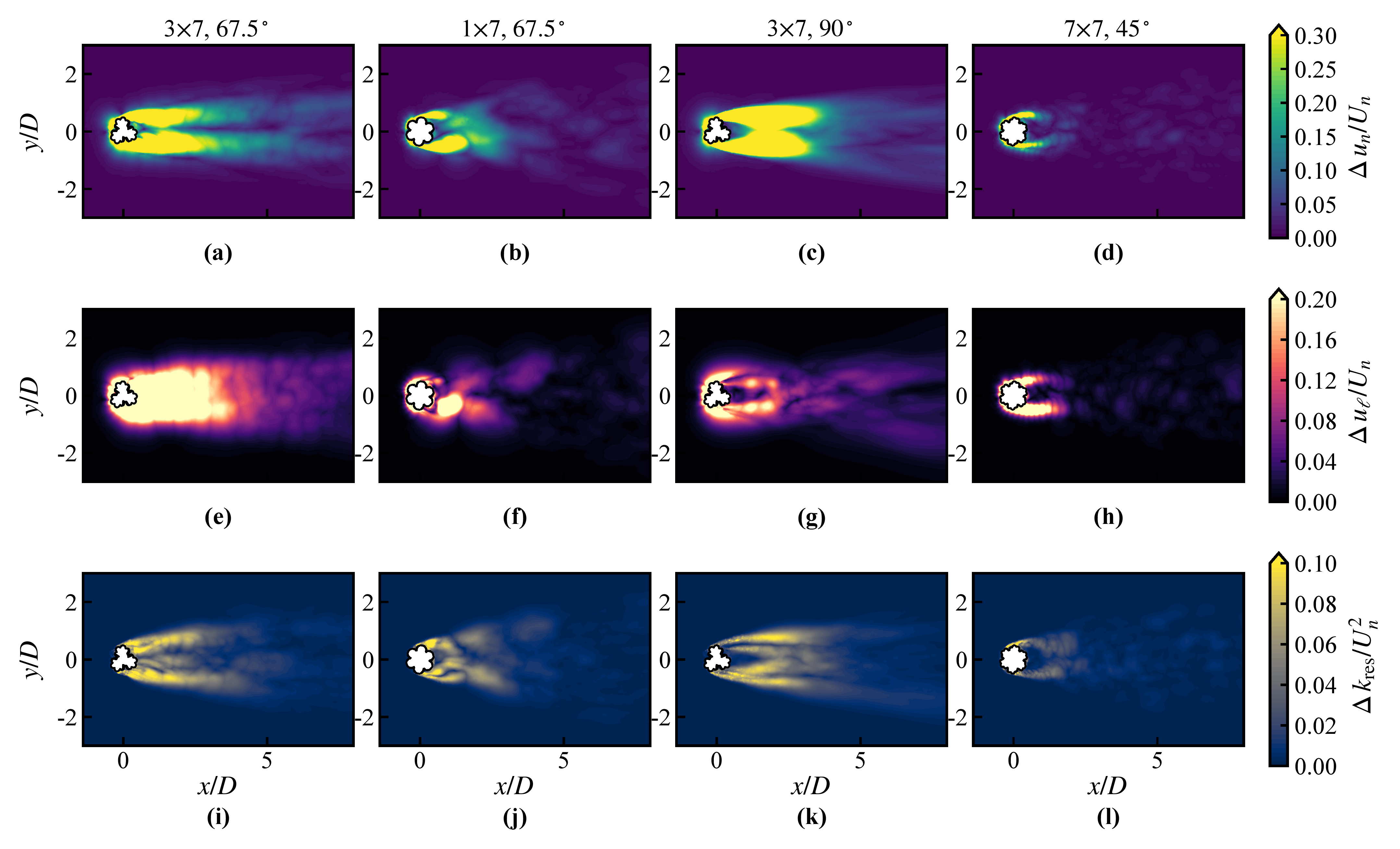}
\caption{Axial phase-to-phase spread distributions, \(\Delta Q(x,y)=\max_p Q_p(x,y)-\min_p Q_p(x,y)\), across four representative stranded cable cases: (a)--(d) normal velocity spread \(\Delta u_n/U_n\); (e)--(h) lateral velocity spread \(\Delta u_\ell/U_n\); and (i)--(l) resolved turbulent kinetic energy spread \(\Delta k_{\mathrm{res}}/U_n^2\).}
\label{fig:fig13}
\end{figure*}

Figures~\ref{fig:fig11}(a)--\ref{fig:fig11}(f) show the crossflow distribution of the turbulent fluctuations associated with these global loads. At $\alpha = 67.5^\circ$, the lateral Reynolds stress profiles ($R_{\ell\ell}/U_n^2$) for the $3\times7$ cable form broadened, asymmetric lobes whose peak values and positions change across axial sections: peak values remain between $0.094$ and $0.130$, with their transverse locations shifting between $y/D = -0.52$ and $0.51$. The $1\times19$ cable, however, maintains a sharp, centrally concentrated peak near $y/D = 0$, reaching $R_{\ell\ell}/U_n^2 = 0.406$--$0.454$. This comparison relates the mean lateral force and the lateral-force fluctuation observed in Sec.~\ref{sec:global-loads} to the wake turbulence. For the $1\times19$ cable, turbulent fluctuations are focused along the wake centerline, reflecting organized vortex shedding that drives a strong lateral-force fluctuation ($C_{\ell,\mathrm{rms}}(q_{\mathrm{tot}}) = 0.300$) without a steady sideways bias. For the $3\times7$ cable, the helical grooves steer the wake sideways and disperse fluctuation energy across a broader transverse region, and the lateral-force fluctuation is lower ($C_{\ell,\mathrm{rms}}(q_{\mathrm{tot}}) = 0.053$). At $\alpha = 45^\circ$, the lateral fluctuation peaks of the $3\times7$ cable remain on the negative-$y$ side ($y/D = -0.36$ to $-0.23$). The tangential normal-stress profiles ($R_{tt}/U_n^2$, Figs.~\ref{fig:fig11}(d)--\ref{fig:fig11}(f)) likewise show section-dependent shifts in peak position and magnitude, indicating that the helical surface redistributes the directional turbulence energy in space.

To reveal the spatial origin of this axial variability, the $3\times7$ cable at $\alpha = 67.5^\circ$ is visualized in Fig.~\ref{fig:fig12} as the representative case combining the largest mean lateral force, maximum transverse wake deflection, and widest variation in recirculation length. Across its six axial phase planes, the low-velocity deficit core continuously alters its shape and cross-stream position in response to the rotating strand--groove exterior contour. Concurrently, localized regions of positive and negative lateral velocity shift their positions and extent within the near wake ($x/D \le 3$). When viewed alongside the Reynolds stress profiles in Fig.~\ref{fig:fig11}, these two-dimensional velocity fields show that the asymmetric mean lateral force is reflected in the wake as a three-dimensional restructuring of the deficit core and transverse flow, which is closely linked to the helical boundary-layer separation lines.

Complementing the scalar ranges in Fig.~\ref{fig:fig10}, the spatial phase spread locates section-to-section differences in the crossflow plane and shows how far they persist downstream. It is defined as
\begin{equation}
\Delta Q(x,y)=\max_p Q_p(x,y)-\min_p Q_p(x,y),
\label{eq:wake-phase-spread}
\end{equation}
where $p$ indexes the axial sections and $Q_p(x,y)$ represents the time-averaged velocity or resolved turbulent kinetic energy at each crossflow position. As shown in Fig.~\ref{fig:fig13} for the four representative stranded cases, the lateral velocity spread ($\Delta u_\ell/U_n$) for the $3\times7$ cable at $\alpha = 67.5^\circ$ is large throughout the near wake and exceeds that of the single-strand $1\times7$ cable. Under pure crossflow, the $3\times7$ cable develops a broader normal velocity spread ($\Delta u_n/U_n$), and the $7\times7$ cable at $\alpha = 45^\circ$ shows the smallest overall variation among the cases examined. Across all configurations, the velocity and resolved turbulent kinetic energy spreads decay rapidly with downstream distance: the geometric-phase imprint is concentrated in the near-wake recirculation and shear-layer zones ($x/D \lesssim 3$), although for the $3\times7$ cable a normal velocity spread of order $0.1U_n$ persists to $x/D \approx 5$; farther downstream, the imprint is largely smoothed out by turbulent mixing.

In summary, the wake characteristics identified in Sec.~\ref{sec:wake-organization} link the near-wall surface patterns to the differences in global loads. The prominent mean lateral force on the $3\times7$ cable is accompanied by transverse wake deflection and localized distortions of the recirculation zone associated with asymmetric boundary-layer separation, while the transverse dispersal of turbulent fluctuations coincides with a weak lateral-force fluctuation. The $1\times19$ cable, by contrast, shows no mean-wake deflection; its symmetric mean wake concentrates unsteady energy along the centerline, and its lateral-force fluctuation is strong. This comparison shows that the decoupling between the mean lateral force and the lateral-force fluctuation is reflected in the three-dimensional organization of the near wake. However, time-averaged wake profiles alone cannot reveal whether local shedding frequencies align with the global force response. Section~\ref{sec:unsteady-force} therefore investigates the time-dependent aerodynamic coupling, utilizing lateral-force spectra and downstream force--probe coherence to determine how wake vortex dynamics relate to global unsteady loading.

\subsection{Unsteady force signatures and wake--force coupling}
\label{sec:unsteady-force}

\begingroup
% Keep the existing natural display gaps, without column-fill stretching.
\abovedisplayskip=\dimexpr\abovedisplayskip\relax
\belowdisplayskip=\dimexpr\belowdisplayskip\relax
\abovedisplayshortskip=\dimexpr\abovedisplayshortskip\relax
\belowdisplayshortskip=\dimexpr\belowdisplayshortskip\relax

The nominal twelve-wire outer layer of the single-strand \(1\times19\) cable provides a wire-scale surface topography; the tri-lobed, three-strand \(3\times7\) cable instead combines the helical lay angles of both the outer strands and the internal wires. The contrasting surface pressure, separation patterns, and wake topologies identified at \(67.5^\circ\) in Secs.~\ref{sec:surface-organization} and~\ref{sec:wake-organization} motivate a direct comparison of lateral-force fluctuation amplitudes, spectral characteristics, and local force--velocity coherence. Subsequent diagnostics focus primarily on the lateral force, where the decoupling between the mean lateral force and the lateral-force fluctuation is most evident; normal and tangential loads remain part of the three-component balance. The unsteady lateral-force coefficient and its centered root-mean-square fluctuation are defined as
\begin{equation}
\begin{gathered}
C_\ell(t)=\frac{F_\ell(t)}{q_n A_{\mathrm{ref}}},\qquad
C_\ell=\langle C_\ell(t)\rangle,\\
C_{\ell,\mathrm{rms}}
=\sqrt{\langle[C_\ell(t)-C_\ell]^2\rangle}.
\end{gathered}
\label{eq:sec34-rms}
\end{equation}
Here \(F_\ell(t)\) is the instantaneous lateral force, angle brackets denote the time-averaging operator over the sampling window, and \(C_{\ell,\mathrm{rms}}\) quantifies the overall intensity of unsteady force fluctuations around the mean. Because these coefficients are referenced to \(q_n\), their values are labeled \(C_\ell(q_n)\) and \(C_{\ell,\mathrm{rms}}(q_n)\) in Secs.~\ref{sec:unsteady-force} and \ref{sec:modal-organization}, as distinguished from the \(q_{\mathrm{tot}}\)-normalized values in Secs.~\ref{sec:global-loads}--\ref{sec:wake-organization}.

The dominant vortex shedding frequency and common-band normalized spectral distribution are defined as
\begin{equation}
\begin{gathered}
St_{\mathrm{p}}=\frac{f_{\mathrm{p}}D}{U_n},\\
\widehat S_\ell(St_n)
=\frac{S_\ell(St_n)}
{\displaystyle\int_{\mathcal B}S_\ell(s)\,\mathrm ds}.
\end{gathered}
\label{eq:sec34-shape}
\end{equation}
Here \(St_n\) is the normal Strouhal number of Eq.~\eqref{eq:dimensionless} evaluated at frequency \(f\), \(\mathcal B=[0.05,1.0]\) is the evaluation band, \(f_{\mathrm{p}}\) denotes the frequency of the force power spectral density (PSD) maximum for \(0.1\le St_n\le1.0\), which excludes the subharmonic modulation of the recirculation region, and \(S_\ell(St_n)\) is the one-sided PSD per unit \(St_n\), with \(s\) serving as the integration dummy variable. Power spectral densities are evaluated using Welch's method\cite{Welch1967} applied to synchronized force time histories, with Hann-windowed segments of 16\,384 samples, 50\% overlap, and linear segment detrending (8192-sample segments for the coherence estimates below); the resulting frequency resolution is \(\Delta St_n \approx 0.013\text{--}0.035\), depending on inclination. Converting the spectra from physical frequency (Hz) referenced to total dynamic pressure \(q_{\mathrm{tot}}\) yields
\begin{equation}
S_\ell(St_n)=\frac{U_n}{D}
\left(\frac{q_{\mathrm{tot}}}{q_n}\right)^2
\times S_{\ell,q_{\mathrm{tot}}}^{f}
\left(\frac{U_n}{D}St_n\right),
\label{eq:sec34-density}
\end{equation}
where \(q_{\mathrm{tot}}\) denotes the total dynamic pressure and superscript \(f\) designates the one-sided PSD per unit Hz. This conversion accounts for dynamic pressure rescaling and the frequency Jacobian (\(\mathrm{d}f/\mathrm{d}St_n = U_n/D\)). Normalizing by the integrated area over the common frequency band decouples the spectral distribution from the absolute fluctuation amplitude, with band integrals computed using piecewise-linear interpolation between discrete frequency bins.

\begin{figure*}[tp]
\centering
\setlength{\abovecaptionskip}{5pt}
\includegraphics[width=0.557\textwidth]{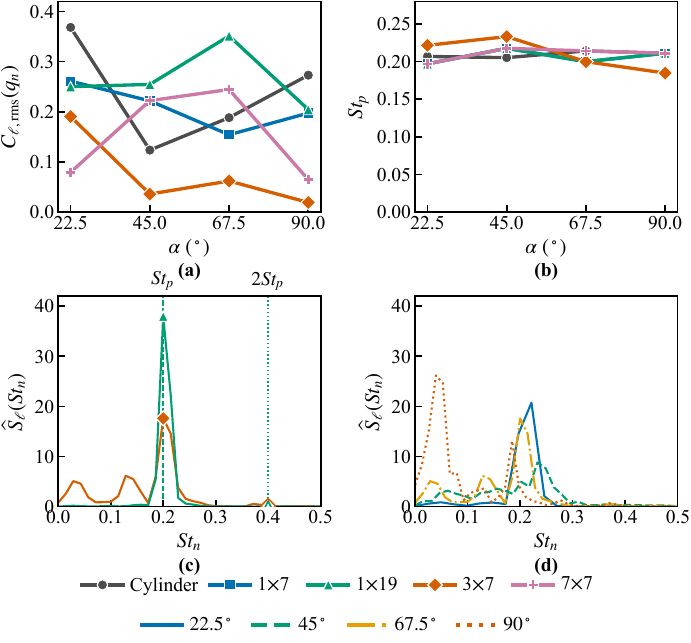}
\caption{Unsteady lateral aerodynamic forces and spectral characteristics across cable configurations and inclinations: (a) root-mean-square lateral-force coefficient \(C_{\ell,\mathrm{rms}}(q_n)\); (b) dominant vortex shedding frequency (peak Strouhal number) \(St_{\mathrm{p}}\); (c) area-normalized lateral-force spectra \(\widehat{S}_\ell(St_n)\) comparing \(3\times7\) and \(1\times19\) cables at \(\alpha = 67.5^\circ\); and (d) spectral evolution of the \(3\times7\) cable across all four inclinations \(\alpha = 22.5^\circ\text{--}90^\circ\).}
\label{fig:unsteady-force-matrix}
\end{figure*}

\begin{figure*}[tp]
\centering
\setlength{\abovecaptionskip}{5pt}
\includegraphics[width=0.554\textwidth]{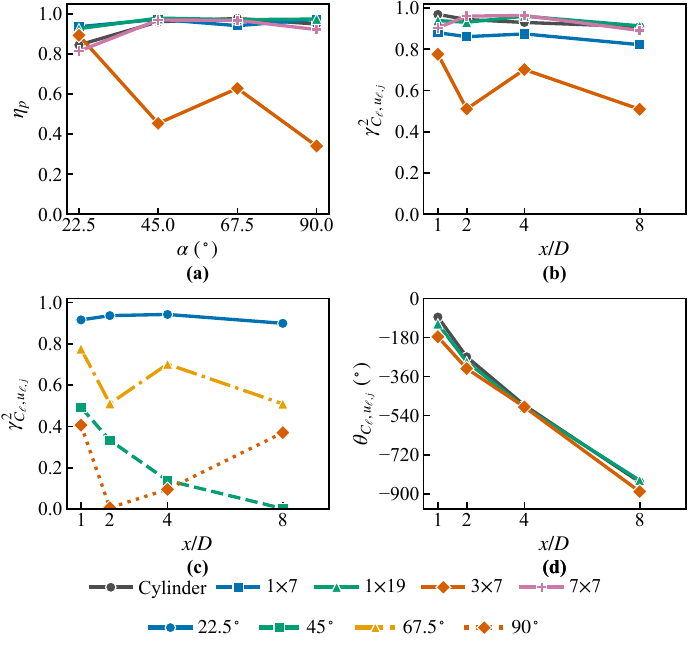}
\caption{Spectral energy concentration and downstream force--wake association: (a) peak-neighborhood energy fraction \(\eta_{\mathrm{p}}\) across inflow inclinations; (b) downstream magnitude-squared coherence \(\gamma^2_{C_\ell,u_{\ell,j}}\) across all five geometries at \(\alpha = 67.5^\circ\); (c) downstream coherence evolution of the \(3\times7\) cable across all four inclinations; and (d) unwrapped downstream probe phase lag \(\theta_{C_\ell,u_{\ell,j}}\) along the wake centerline for the smooth cylinder, \(1\times19\), and \(3\times7\) cables at \(\alpha = 67.5^\circ\).}
\label{fig:force-wake-coherence-phase}
\end{figure*}

Figure~\ref{fig:unsteady-force-matrix} summarizes the unsteady lateral forces and spectral distributions across configurations and inclinations. For the smooth cylinder, a clear shedding peak is maintained near \(St_n\approx0.205\text{--}0.214\) across all inclination angles [Fig.~\ref{fig:unsteady-force-matrix}(b)], reflecting stable quasi-periodic vortex shedding, consistent with the independence principle for yawed cylinders.\cite{Zhao2009YawedCylinder,Lou2022YawedSquare} A similarly concentrated narrow-band response is observed for the \(1\times7\) and \(1\times19\) cables, where the primary frequencies remain within \(St_n\approx0.197\text{--}0.218\). In contrast, the spectrum of the \(3\times7\) cable is more sensitive to inclination, whereas the \(7\times7\) cable departs from narrow-band behavior only at \(22.5^\circ\) (\(\eta_{\mathrm{p}} = 0.81\)) [Fig.~\ref{fig:force-wake-coherence-phase}(a)]. Although \(3\times7\) maintains a prominent primary peak at \(22.5^\circ\) (\(St_n\approx0.222\)), its spectral response broadens considerably at higher inclinations (\(45^\circ\text{--}90^\circ\)) [Fig.~\ref{fig:unsteady-force-matrix}(d)]. In particular, at \(90^\circ\), where the lateral-force fluctuation is suppressed to \(C_{\ell,\mathrm{rms}}(q_n)=0.019\) [Fig.~\ref{fig:unsteady-force-matrix}(a)], the shedding peak at \(St_n\approx0.185\) becomes weaker than the broadband low-frequency content at \(St_n\approx0.03\text{--}0.05\), which reflects slow modulation of the elongated recirculation region. Centerline probes beyond the recirculation zone (\(x/D=4\) and 8) confirm shedding at \(St_n\approx0.20\). This broadband redistribution coincides with a substantial reduction in force fluctuations; for instance, \(C_{\ell,\mathrm{rms}}(q_n)\) for \(3\times7\) decreases from 0.190 at \(22.5^\circ\) to 0.036 at \(45^\circ\) [Fig.~\ref{fig:unsteady-force-matrix}(a)].

To measure spectral concentration and relative harmonic content, the peak-neighborhood energy fraction \(\eta_{\mathrm{p}}\) and second-harmonic ratio \(H_2\) are defined as
\begin{equation}
\eta_{\mathrm{p}}=
\frac{\displaystyle\int_{\mathcal B_{\mathrm{p}}}S_\ell(s)\,\mathrm ds}
{\displaystyle\int_{\mathcal B}S_\ell(s)\,\mathrm ds},
\qquad
H_2=\frac{S_\ell(2St_{\mathrm{p}})}{S_\ell(St_{\mathrm{p}})},
\label{eq:sec34-diagnostics}
\end{equation}
where \(\mathcal B_{\mathrm{p}}=\mathcal B\cap[St_{\mathrm{p}}-0.04,St_{\mathrm{p}}+0.04]\) with the fixed half-width 0.04 defining peak-neighborhood power within \(\mathcal B\), and \(H_2\) compares PSD point values evaluated via linear interpolation. For the representative cases in Fig.~\ref{fig:force-wake-coherence-phase}(a) at \(67.5^\circ\), the peak-neighborhood energy fraction \(\eta_{\mathrm{p}}\) reaches 0.971 for \(1\times19\), compared with only 0.628 for \(3\times7\). This spectral variation is closely associated with the wake organization: the regular, spanwise-coherent shedding of \(1\times19\) concentrates energy around the fundamental frequency, whereas the three-dimensional shear-layer interactions around \(3\times7\) distribute fluctuations across adjacent bands and exhibit greater relative energy at the second harmonic (\(H_2 = 6.11\times10^{-3}\) versus \(2.11\times10^{-4}\)).

The frequency-specific linear association of global lateral force with the lateral velocity \(u_{\ell,j}\) at downstream probe \(j\) is quantified by magnitude-squared coherence,\cite{Carter1987Coherence}
\begin{equation}
\gamma^2_{C_\ell,u_{\ell,j}}(f)=
\frac{\left|S_{C_\ell u_{\ell,j}}^{(\mathrm{coh})}(f)\right|^2}
{S_{C_\ell C_\ell}^{(\mathrm{coh})}(f)
 S_{u_{\ell,j}u_{\ell,j}}^{(\mathrm{coh})}(f)},
\label{eq:sec34-msc}
\end{equation}
where \(S_{C_\ell u_{\ell,j}}^{(\mathrm{coh})}(f)\) is the cross-spectral density, and \(S_{C_\ell C_\ell}^{(\mathrm{coh})}(f)\) and \(S_{u_{\ell,j}u_{\ell,j}}^{(\mathrm{coh})}(f)\) are the autospectral densities evaluated from synchronized time segments. On the discrete coherence frequency grid \(f_k^{(\mathrm{coh})}\), the representative shedding frequency \(f_{\mathrm{c}}\) nearest to the force spectral peak \(f_{\mathrm{p}}\) and the corresponding cross-spectral phase lag \(\theta_{C_\ell,u_{\ell,j}}\) are identified via
\begin{equation}
\begin{gathered}
k_{\mathrm{c}}=\mathop{\mathrm{arg\,min}}_k |f_k^{(\mathrm{coh})}-f_{\mathrm{p}}|,\qquad
f_{\mathrm{c}}=f_{k_{\mathrm{c}}}^{(\mathrm{coh})},\\
\theta_{C_\ell,u_{\ell,j}}(f_{\mathrm{c}}) = \arg S_{C_\ell u_{\ell,j}}^{(\mathrm{coh})}(f_{\mathrm{c}}).
\end{gathered}
\label{eq:sec34-phase}
\end{equation}
Here, \(\theta_{C_\ell,u_{\ell,j}}\) designates the phase lag of local wake velocity fluctuations relative to the global lateral load (negative when \(u_{\ell,j}\) lags \(C_\ell\)), as distinguished from the axial geometric phase \(\phi\) defined in Eq.~\eqref{eq:phase}. To continuously monitor the spatial advection of shedding structures, the phase is unwrapped along the downstream probe array (\(x/D = 1\text{--}8\)), with an operational threshold of \(\gamma^2(f_{\mathrm{c}}) \ge 0.5\) enforced for physical phase interpretation.

The resulting force--wake coherence diagnostics in Fig.~\ref{fig:force-wake-coherence-phase} indicate that the spatial organization of the wake depends on the exterior surface morphology and its orientation to the inflow. As shown in Fig.~\ref{fig:force-wake-coherence-phase}(a), the smooth cylinder and single-strand cables (\(1\times7\), \(1\times19\)) maintain lateral fluctuation energy tightly concentrated within the primary shedding band (\(\eta_{\mathrm{p}} \approx 0.85\text{--}0.98\)) across all inclinations. At \(\alpha = 67.5^\circ\), this high spectral concentration is accompanied by persistent downstream linear coupling [Fig.~\ref{fig:force-wake-coherence-phase}(b)], with \(1\times19\) exhibiting high coherence (\(\gamma^2 = 0.914\text{--}0.960\)) throughout \(x/D=1\text{--}8\), corresponding to regular, spanwise-coherent K\'arm\'an vortices advecting downstream with minimal structural deformation. In contrast, the multi-strand \(3\times7\) cable exhibits marked spatial decorrelation: coherence falls to 0.511 at \(x/D=2\) and remains degraded into the intermediate wake (0.509 at \(x/D=8\)), consistent with the spectral broadening identified in Fig.~\ref{fig:force-wake-coherence-phase}(a), where \(\eta_{\mathrm{p}}\) drops to 0.628. This loss of coherence depends strongly on the inclination [Fig.~\ref{fig:force-wake-coherence-phase}(c)]: the \(22.5^\circ\) case retains strong wake association (\(\gamma^2 = 0.900\text{--}0.943\)), while the \(45^\circ\) and \(90^\circ\) cases drop below the interpretive threshold across the near wake, indicating that the integrated lateral force is only weakly coupled to the local wake velocity at these inclinations. At \(90^\circ\), however, the centerline probes at \(x/D=4\) and 8 still detect shedding at \(St_n\approx0.20\).

Along the wake centerline [Fig.~\ref{fig:force-wake-coherence-phase}(d)], the unwrapped phase trajectories for the smooth cylinder, \(1\times19\), and \(3\times7\) cables decrease at an approximately uniform rate beyond \(x/D = 2\) (\(\mathrm{d}\theta/\mathrm{d}(x/D)\) between \(-91^\circ\) and \(-96^\circ\) over \(2 \le x/D \le 8\)), characteristic of stable vortex convection at \(U_{\mathrm{c}} \approx 0.75\text{--}0.79\,U_n\). The phase curves of the \(1\times19\) and \(3\times7\) cables intersect near \(-500^\circ\) at \(x/D=4\) and differ by \(\Delta\theta \approx 52^\circ\) at \(x/D=8\), which may reflect a slight convective delay and delayed vortex formation associated with the helical strand grooves.

In summary, the unsteady force and coherence diagnostics in Sec.~\ref{sec:unsteady-force} show two distinct aerodynamic responses associated with cable exterior morphology under moderate-to-high oblique inflow. The single-strand \(1\times19\) construction retains a quasi-circular contour dominated by circumferentially regular, wire-scale textures. This preserves spanwise-coherent K\'arm\'an vortex shedding, which concentrates unsteady lateral energy within a narrow primary band (\(\eta_{\mathrm{p}} \approx 0.97\)) and yields a large lateral-force fluctuation (\(C_{\ell,\mathrm{rms}}(q_n) \approx 0.35\)) and persistent downstream force--wake coherence, while the mean lateral force remains small. The single-strand \(1\times7\) cable likewise retains coherent shedding, but carries an appreciable mean lateral force at intermediate inclinations. In contrast, the multi-strand \(3\times7\) geometry introduces deep helical strand-scale recesses that induce asymmetric separation and lateral wake deflection at intermediate inclinations (\(\alpha = 45^\circ\text{--}67.5^\circ\)). The resulting three-dimensional shear-layer interactions disrupt spanwise vortex synchronization, redistribute fluctuation energy across adjacent frequencies, and reduce the lateral-force fluctuation (\(C_{\ell,\mathrm{rms}}(q_n) \approx 0.06\)), while a mean lateral force is maintained. These results show that the vortex-induced lateral-force fluctuation and the mean lateral force are decoupled in stranded cable aeromechanics. To resolve the spatial coherent structures governing these distinct spectral signatures at the primary shedding frequency, Sec.~\ref{sec:modal-organization} subsequently examines the modal organization of the wake using SPOD.

\par
\endgroup

\subsection{Modal organization at the force-nearest wake frequency}
\label{sec:modal-organization}

The preceding analyses show that the nominal exterior geometry and inflow inclination jointly control surface-pressure distributions, boundary-layer separation lines, and the spectral concentration of unsteady lateral forces. A key unresolved question is how these contrasting force signatures are manifested in the spatial organization of the downstream wake at the primary vortex shedding frequency. As in the preceding sections, the analysis focuses on the representative contrast between the \(1\times19\) and \(3\times7\) cables at \(\alpha = 67.5^\circ\). Spectral proper orthogonal decomposition (SPOD)\cite{Towne2018,Schmidt2020SPODGuide} is employed as an energy-optimal, frequency-resolved diagnostic. By targeting the discrete SPOD frequency closest to the force spectral peak \(f_{\mathrm{p}}\), the analysis extracts the dominant spatial coherent structures within the wake and evaluates their streamwise and axial coherence.

\begin{figure*}[tp]
\centering
\setlength{\abovecaptionskip}{5pt}
\includegraphics[width=0.553\textwidth]{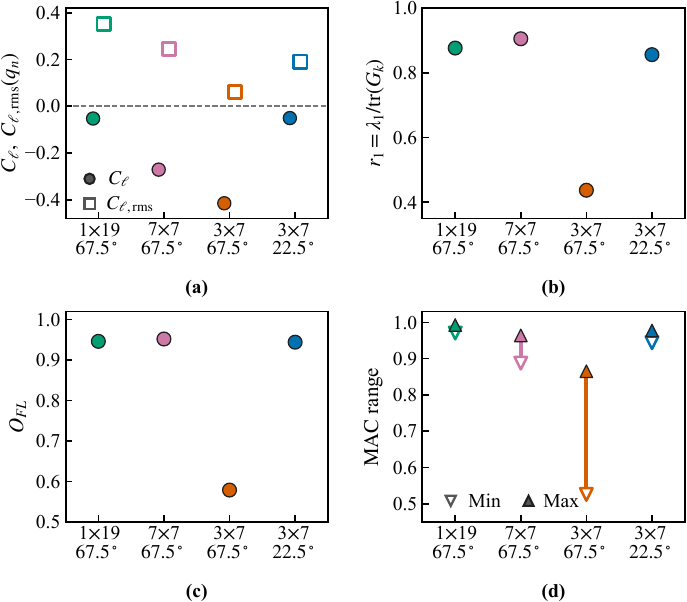}
\caption{Modal diagnostics and aerodynamic force characteristics across representative cases at the primary wake frequency: (a) mean lateral-force coefficient \(C_\ell(q_n)\) and centered root-mean-square \(C_{\ell,\mathrm{rms}}(q_n)\); (b) leading SPOD modal energy fraction \(r_1\); (c) first-half/second-half modal overlap \(O_{\mathrm{FL}}\); and (d) spanwise phase-MAC variation range across axial sections.}
\label{fig:modal-summary}
\end{figure*}

For each cross-sectional snapshot, the velocity fluctuation vector in the cable-fixed frame is denoted by \(\boldsymbol{v}' = [\boldsymbol{u}_n', \boldsymbol{u}_\ell', \boldsymbol{u}_t']^{\mathrm{T}}\). To account for the local mesh resolution, the area-weighted inner product is defined through the diagonal matrix \(W = \operatorname{diag}(a_1, \ldots, a_N, a_1, \ldots, a_N, a_1, \ldots, a_N)\), where \(a_j\) represents the area of fluid cell \(j\). Following Welch's averaged-periodogram approach\cite{Welch1967} as adapted for SPOD,\cite{Towne2018} the velocity fluctuations, weighted by \(W^{1/2}\), are segmented into Hann-windowed blocks with 50\% overlap to construct the cross-spectral density (CSD) matrix \(G_k\) at discrete frequency \(f_k\). The energy-optimal coherent structures are then extracted via the Fredholm eigenvalue problem:
\begin{equation}
G_k \Phi_k = \Phi_k \Lambda_k, \qquad
r_1(k) = \frac{\lambda_1(G_k)}{\operatorname{tr}(G_k)},
\label{eq:section35-spod}
\end{equation}
where \(\Lambda_k = \operatorname{diag}(\lambda_1, \lambda_2, \ldots)\) contains the modal energies in descending order (\(\lambda_1 \ge \lambda_2 \ge \cdots\)), and \(r_1\) represents the leading modal energy fraction quantifying the degree of low-rank spectral organization. The analysis isolates the discrete frequency bin \(f_{\mathrm{SPOD}}\) closest to the force spectral peak \(f_{\mathrm{p}}\). The reconstructed physical mode \(\boldsymbol{\psi}\), the leading column of \(W^{-1/2}\Phi_k\), satisfies the spatial normalization \(\sum_j a_j |\boldsymbol{\psi}_j|^2 = 1\), so that \(\boldsymbol{\psi}\) has units of \(\mathrm{m}^{-1}\); consequently, the modal fields depict normalized wavepacket topology and phase distributions, independent of the absolute fluctuation amplitude.

\looseness=1\relax To assess the statistical convergence and spatial uniformity of the coherent structures, two metrics are evaluated. The temporal convergence of the leading SPOD mode across the sampling record is quantified by the first-half/second-half modal overlap:
\begin{equation}
O_{\mathrm{FL}}=
\frac{\left|\boldsymbol{\psi}_{\mathrm{F}}^{H}W\boldsymbol{\psi}_{\mathrm{L}}\right|}
{\left[(\boldsymbol{\psi}_{\mathrm{F}}^{H}W\boldsymbol{\psi}_{\mathrm{F}})(\boldsymbol{\psi}_{\mathrm{L}}^{H}W\boldsymbol{\psi}_{\mathrm{L}})\right]^{1/2}},
\label{eq:section35-ofl}
\end{equation}
where \(\boldsymbol{\psi}_{\mathrm{F}}\) and \(\boldsymbol{\psi}_{\mathrm{L}}\) denote the leading modes computed from the first and second halves of the Welch block ensemble, respectively. Values approaching unity indicate that the extracted mode represents a statistically stationary, converged coherent wake mode over the sampling record. Furthermore, to quantify the structural invariance along the cable axis, the modal similarity between different axial cross sections \(p\) and \(q\) is evaluated via the modal assurance criterion (MAC):
\begin{equation}
\operatorname{MAC}_{pq}=
\frac{\left|\boldsymbol{\psi}_p^{H}W_{\mathrm{c}}\boldsymbol{\psi}_q\right|^2}
{(\boldsymbol{\psi}_p^{H}W_{\mathrm{c}}\boldsymbol{\psi}_p)(\boldsymbol{\psi}_q^{H}W_{\mathrm{c}}\boldsymbol{\psi}_q)},
\label{eq:section35-mac}
\end{equation}
evaluated over the common fluid domain with area weighting \(W_{\mathrm{c}}\). The resulting \(\mathrm{MAC}\) range (\([0, 1]\)) provides a metric of spanwise coherence: values near unity denote an axially invariant shedding topology, whereas lower values indicate spatial distortion and phase decorrelation along the span.

\begin{figure*}[tp]
\centering
\setlength{\abovecaptionskip}{5pt}
\includegraphics[width=0.90\textwidth]{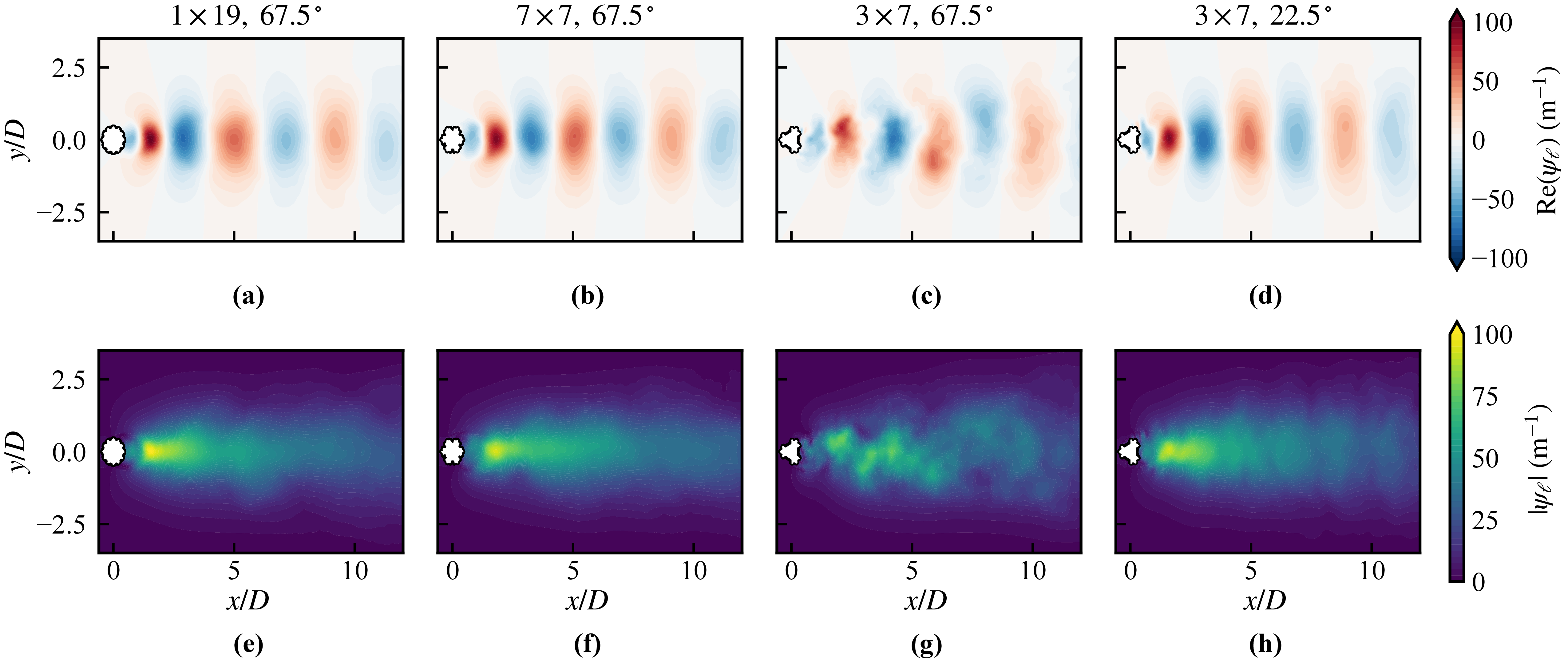}
\caption{Leading SPOD lateral modal fields at the primary wake frequency across representative cases: (a)--(d) real part \(\operatorname{Re}(\psi_\ell)\), showing the phase-aligned coherent structures; (e)--(h) modal envelope magnitude \(\lvert\psi_\ell\rvert\), showing the distribution of normalized modal amplitude. The black outline traces the actual cable surface envelope.}
\label{fig:representative-modal-fields}
\end{figure*}

\looseness=1\relax The primary construction contrast is evaluated at \(\alpha = 67.5^\circ\). For the \(1 \times 19\) cable, the force-nearest SPOD frequency yields a leading energy fraction of \(r_1 = 0.876\) [Fig.~\ref{fig:modal-summary}(b)], a temporal overlap of \(O_{\mathrm{FL}} = 0.946\) [Fig.~\ref{fig:modal-summary}(c)], and an axial similarity range of \(\mathrm{MAC} \in [0.971, 0.992]\) [Fig.~\ref{fig:modal-summary}(d)]. A comparable degree of modal coherence is obtained for the \(7 \times 7\) cable at the same inclination, yielding \(r_1 = 0.905\), \(O_{\mathrm{FL}} = 0.952\), and \(\mathrm{MAC} \in [0.888, 0.964]\). Compared with these two cables, the \(3 \times 7\) cable exhibits marked spectral dispersion and spatial decorrelation, registering \(r_1 = 0.437\), \(O_{\mathrm{FL}} = 0.579\), and \(\mathrm{MAC} \in [0.526, 0.865]\). The spatial modal fields in Fig.~\ref{fig:representative-modal-fields} reflect this quantitative difference: the \(1 \times 19\) and \(7 \times 7\) cables display compact, antisymmetrically alternating vortex cores [Figs.~\ref{fig:representative-modal-fields}(a) and \ref{fig:representative-modal-fields}(b)] with focused near-wake contours [Figs.~\ref{fig:representative-modal-fields}(e) and \ref{fig:representative-modal-fields}(f)]. In contrast, the \(3 \times 7\) cable develops a broadened, asymmetric mode whose envelope lies slightly on the positive lateral side near \(x/D \approx 2\) and shifts toward the negative lateral side farther downstream [Figs.~\ref{fig:representative-modal-fields}(c) and \ref{fig:representative-modal-fields}(g)], accompanied by its large mean lateral force (\(C_\ell(q_n) = -0.415\)) and suppressed fluctuations [\(C_{\ell,\mathrm{rms}}(q_n) = 0.062\), Fig.~\ref{fig:modal-summary}(a)]. When the inclination is reduced to \(\alpha = 22.5^\circ\), the \(3 \times 7\) configuration recovers dominant low-rank organization (\(r_1 = 0.856\), \(O_{\mathrm{FL}} = 0.944\), \(\mathrm{MAC} \in [0.944, 0.977]\)), restoring symmetric alternating shedding and a concentrated focal contour [Figs.~\ref{fig:representative-modal-fields}(d) and \ref{fig:representative-modal-fields}(h)], and the mean lateral-force magnitude decreases to \(\lvert C_\ell(q_n)\rvert = 0.051\). This recovery indicates that modal decoherence and steady lateral deflection are associated with the deep strand-scale grooves at intermediate inclinations.

\looseness=1\relax These modal diagnostics complete the chain linking the exterior surface geometry to the aerodynamic loads. For the single-strand \(1\times19\) cable, the quasi-circular contour with high circumferential symmetry confines boundary-layer separation to the lateral sides, maintaining spanwise-synchronized K\'arm\'an vortex shedding; this concentrates fluctuation energy into a dominant low-rank coherent wake mode, yielding a strong lateral-force fluctuation with a minimal mean lateral force. On the other hand, the tri-lobed contour of the \(3\times7\) cable features a threefold symmetry with strand-scale helical grooves that channel the crossflow and induce asymmetric local separation and lateral crossflow deflection at intermediate inclinations (\(\alpha = 45^\circ\text{--}67.5^\circ\)). The resulting shear-layer distortion disrupts spanwise vortex synchronization and spreads fluctuation energy across higher modes, and a mean lateral force is sustained. This interpretation is consistent with the behavior of the \(7\times7\) cable, whose regular hexagonal envelope provides sufficient symmetry to maintain coherent vortex shedding while the cable carries an appreciable mean lateral force. It is also consistent with the recovery of shedding coherence in the \(3\times7\) cable at the small inclination \(\alpha = 22.5^\circ\), where crossflow deflection is diminished. The aeromechanics of stranded cables under oblique inflow is thus associated mainly with cross-sectional symmetry and groove depth. By resolving how coherent wake modes either preserve spanwise synchronization or undergo lateral deflection, the force-nearest SPOD modes provide a spatial link between local surface flow deflection and the directional aerodynamic response of inclined towing cables.

\section{Conclusions}

Large-eddy simulations of a smooth cylinder and four stranded cable constructions across four inflow inclinations at $Re\approx1.0\times10^4$ indicate that exterior surface morphology and flow orientation jointly determine the three-dimensional aerodynamic response through coupled surface separation, wake reorganization, and direction-specific loading. The differences between the cable geometries are associated mainly with the cross-sectional circumferential symmetry order and the strand-scale groove depth. Cables featuring highly symmetric, quasi-circular outer envelopes and fine wire-scale surface textures (represented by the $1\times19$ construction) promote balanced crossflow and regular vortex shedding. Configurations characterized by low discrete symmetry and deep strand-scale helical grooves (represented by the $3\times7$ construction), on the other hand, present a tri-lobed cross section that induces clear lateral flow deflection and wake restructuring under oblique inflow.

The cable exterior morphology leads to distinct flow evolutions from the near-wall region into the wake. Under moderate-to-high oblique inclinations, quasi-circular envelopes confine boundary-layer separation symmetrically to the lateral flanks, sustaining spanwise-synchronized K\'arm\'an vortex shedding that concentrates fluctuation energy into a dominant low-rank coherent mode. In contrast, deep strand-scale grooves channel the local crossflow and are accompanied by asymmetric separation-line pinning and pronounced axial phase modulations; these three-dimensional shear interactions disrupt spanwise vortex coherence, deflect the recirculation zone laterally, and disperse fluctuation energy across broadband scales. The retention or breakdown of shedding coherence is linked with the intensity of groove-induced crossflow channeling: the hexagonal $7\times7$ cable preserves coherent shedding owing to its regular circumferential symmetry, while the $3\times7$ cable recovers coherence at small inclinations as crossflow steering diminishes.

These near-wall separation and wake characteristics are reflected in the integrated aerodynamic loads through the surface tractions. Force-source decompositions show that surface pressure accounts for 94.9\%--98.3\% of the mean normal force across all cases and dominates the mean lateral force (98\% for the $3\times7$ cable at $\alpha=67.5^\circ$). Tangential loads remain secondary and arise from a construction-dependent balance between viscous friction and surface pressure. The mean lateral force and the lateral-force fluctuation are decoupled across configurations: quasi-circular geometries maintain balanced separation on opposing flanks, preserving small mean lateral forces while allowing regular vortex shedding to drive periodic lateral-force fluctuations. Deep-grooved configurations, however, lock in asymmetric separation that generates a persistent net transverse pressure difference and an appreciable mean lateral force, while the disruption of shedding coherence attenuates lateral-force fluctuations. The $7\times7$ cable at $\alpha=67.5^\circ$ combines an appreciable mean lateral force with strong fluctuations, so the two quantities vary independently with construction and inclination.

For aerial towed systems employing high-aspect-ratio or long-span stranded cables with stringent towing stability requirements, aerodynamic assessments should not be confined to canonical equivalent-cylinder simplifications, and adopting refined criteria based on paired construction--inclination evaluations is recommended. For fixed rigid cable segments, the mean lateral force and the lateral-force fluctuation are decoupled. A quasi-circular construction such as $1\times19$ carries little mean lateral force but strong periodic fluctuations, whereas a deep-grooved construction such as $3\times7$ carries weak fluctuations but a relatively large mean lateral force. From an aerodynamic standpoint, constructions of the $1\times19$ type therefore suit towing applications that require a small mean lateral force for lateral stability, and constructions of the $3\times7$ type suit applications in which the lateral-force fluctuations transmitted to the towed system at moderate-to-high inclinations must be limited. The $1\times7$ and $7\times7$ constructions combine an appreciable mean lateral force with moderate-to-strong fluctuations at intermediate inclinations. These direction-specific mean and fluctuating loads provide a physical basis for the aerodynamic load models used in multibody dynamics and trajectory simulations of aerial towed systems that account for cable motion and fluid--structure coupling.

Because large-eddy simulations of long cables over many shedding cycles and many configurations remain computationally demanding, the present conclusions are drawn for fixed rigid cable segments at a single freestream Reynolds number, $Re\approx1.0\times10^4$ ($Re_n\approx3.8\times10^3$--$1.0\times10^4$), with a spanwise periodic length of one lay pitch ($6D$), which may not capture the spanwise correlation of much longer cables. Building on the present study, multibody dynamics and trajectory simulations that incorporate the spatial coherence of high-aspect-ratio flexible cables can evaluate the effect of the identified lateral loads on operating towed systems. Wind-tunnel aeroelastic experiments on vortex-induced vibration and fluid--structure interaction at representative inclinations such as $\alpha=67.5^\circ$, together with fatigue tests addressing inter-wire fretting under nonstationary aerodynamic excitation, can further extend these findings.

% ============================================================================================================
% Sections required by AIP Publishing after the Conclusions (no supplementary material, no acknowledgments):
% Author Declarations (Conflict of Interest, Author Contributions) and Data Availability.
% Ethics approval is only required for studies with human participants or animals, so it is omitted.
% In "Author Contributions", replace Author One ... Author Five with the same names as in front_authors.tex.
% CRediT roles (AIP list): Conceptualization; Data curation; Formal analysis; Funding acquisition;
% Investigation; Methodology; Project administration; Resources; Software; Supervision; Validation;
% Visualization; Writing -- original draft; Writing -- review & editing. Degree: lead, equal, or supporting.
% The corresponding author must check this statement, and every author must agree to his or her own roles.
% ============================================================================================================
\section*{Author Declarations}

\subsection*{Conflict of Interest}
The authors have no conflicts to disclose.

\subsection*{Author Contributions}
\textbf{Tielin Ma:} Conceptualization (lead); Supervision (equal); Project administration (equal); Resources (equal); Writing -- review \& editing (supporting).
\textbf{Junlin Zhang:} Methodology (lead); Software (lead); Validation (lead); Formal analysis (lead); Investigation (lead); Data curation (lead); Visualization (lead); Writing -- original draft (lead); Writing -- review \& editing (supporting).
\textbf{Jingcheng Fu:} Conceptualization (supporting); Funding acquisition (lead); Supervision (equal); Project administration (equal); Writing -- review \& editing (lead).
\textbf{Chongwen Jiang:} Resources (equal); Supervision (supporting).
\textbf{Chuan Ye:} Funding acquisition (supporting); Resources (supporting).

\section*{Data Availability}
The data that support the findings of this study are available from the corresponding author upon reasonable request.

% AIP reprint style empties \refname; restore the heading for the reference list
\renewcommand{\refname}{References}
% revtex writes the longbibliography control entry only if \bibliographystyle is not called;
% aipnum4-1pof = aipnum4-1 that also prints the issue number, e.g. Phys. Fluids 35(3), 032101 (2023)
\makeatletter\def\@bibstyle{aipnum4-1pof}\makeatother
\bibliography{references}
\end{document}